\documentclass[preprint,aps,tightenlines,showpacs]{revtex4}
\usepackage[dvips]{graphicx}
\usepackage{dcolumn}
\usepackage{epsfig}
\def\bra#1{\mathinner{\langle{#1}|}}
\def\ket#1{\mathinner{|{#1}\rangle}}
\def\braket#1{\mathinner{\langle{#1}\rangle}}

\begin{document}
\title{Extraction of  Resonances from  
Meson-Nucleon Reactions\footnote{Notice: Authored by
Jefferson Science Associates, LLC under U.S. DOE Contract No. DE-AC05-06OR23177. 
The U.S. Government retains a non-exclusive, paid-up, irrevocable, world-wide
license to publish or reproduce this manuscript for U.S. Government purposes.
} }
\author{N. Suzuki}
\affiliation{ Excited Baryon Analysis Center (EBAC), Thomas Jefferson National
Accelerator Facility, Newport News, VA 23606, USA}
\affiliation{Department of Physics, Osaka University, Toyonaka,
Osaka 560-0043, Japan}
\author{T. Sato}
\affiliation{ Excited Baryon Analysis Center (EBAC), Thomas Jefferson National
Accelerator Facility, Newport News, VA 23606, USA}
\affiliation{Department of Physics, Osaka University, Toyonaka,
Osaka 560-0043, Japan}
\author{T.-S. H. Lee}
\affiliation{ Excited Baryon Analysis Center (EBAC), Thomas Jefferson National
Accelerator Facility, Newport News, VA 23606, USA}
\affiliation{Physics Division, Argonne National Laboratory,
Argonne, IL 60439, USA}

\begin{abstract}

We present a pedagogical study of the 
commonly employed Speed-Plot (SP) and Time-delay (TD) methods
for extracting the  resonance parameters 
from the data of two particle coupled-channels  reactions. 
Within several exactly solvable models,
it is found  that these two methods find  poles on  
different  Riemann sheets and are not always valid.
We then develop an analytic continuation method for extracting
nucleon resonances within a 
dynamical coupled-channel formulation of $\pi N$ and $\gamma N$ reactions.
The main focus of this paper is on resolving the complications due to
the coupling with the unstable $\pi \Delta, \rho N, \sigma N$ channels which
decay into $\pi \pi N$ states. 
By using the results from the considered exactly solvable models,
 explicit numerical procedures are
presented and verified.
As a first application of the
developed analytic continuation method, we present
the nucleon resonances in 
 some partial waves extracted within
a recently developed coupled-channels
model of $\pi N$ reactions.  
The results from this realistic $\pi N$ model, which includes
$\pi N$, $\eta N$, $\pi\Delta$, $\rho N$, and $\sigma N$ channels, also 
show that the simple
pole parametrization of the resonant propagator using the poles
extracted from SP and TD methods works poorly.

\end{abstract}

\maketitle

\newpage

\section{Introduction}

The excited baryon and meson states couple strongly with the continuum
states. Thus they are identified with the
resonance states in  hadron reactions.
The spectra and decay widths of the hadron resonances  reveal
the role of  the confinement and  chiral symmetry of QCD
in  the non-perturbative region. 
Therefore, the extraction of the basic resonance parameters from
reaction data is one of the important tasks in hadron physics.
Ideally, it should involve the following steps:
\begin{enumerate}
\item Perform complete
measurements of all independent observables of the reactions considered.
For example, for pseudo-scalar meson photoproduction
reactions one needs to measure 8 observables\cite{tabakin} : 
differential cross sections,
three single polarizations $\Sigma$, $T$ and $P$, and four double 
polarizations $G$, $H$, $E$, and $F$.
\item Extract the partial-wave amplitudes (PWA) from the data.
Here  we need to solve a non-trivial practical
problem since all observables
are bi-linear combinations of PWA; i.e. $\sigma \sim {f_{L'S'} f^*_{LS}}$. 
\item Extract the resonance parameters from the extracted PWA. Here the often
employed methods are based on the 
Breit-Wigner form\cite{feshbach}, 
Speed-Plot  method of Hoehler\cite{hoehler92,hoehler93}, 
and Time-Delay  method of Wigner\cite{eisenbud,wigner55}. 
A more sophisticated and rigorous method is to use 
the dispersion relations, K-matrix, and dynamical
model to analytically 
continue the PWA to the complex energy plane on which
the resonance poles and residues are determined. 
Extensive works based on 
these three models are reviewed in Ref.\cite{burkertlee}.
\end{enumerate}
In reality, we do not have complete measurements for practically all 
meson-nucleon reactions. Even if the measurements are complete,
the step 2 requires some model assumptions to solve
the inverse bi-linear problem in extracting PWA. This model dependence
must be taken into account in interpreting the extracted
resonance parameters.

In this work, we focus on the step 3 in conjunction with the recent efforts in
extracting the nucleon resonances from very extensive and high quality data
of meson production reactions, as reviewed in Ref.\cite{burkertlee}.
The nucleon resonances $N^*,\Delta$ listed by the Particle Data Group\cite{pdg}
are mainly from 
the analysis of $\pi N$ scattering and pion photoproduction reactions.
The Speed-Plot and Time-Delay methods are most often used in these analyses
since they only require the
PWA determined in the step 2. The purpose of this work is to examine the extent
to which these two methods are valid and to develop an analytic continuation
method within a recently developed dynamical model\cite{msl} of meson 
production reactions in the nucleon resonance region.

\begin{figure}
\begin{center}
\includegraphics[width=10cm]{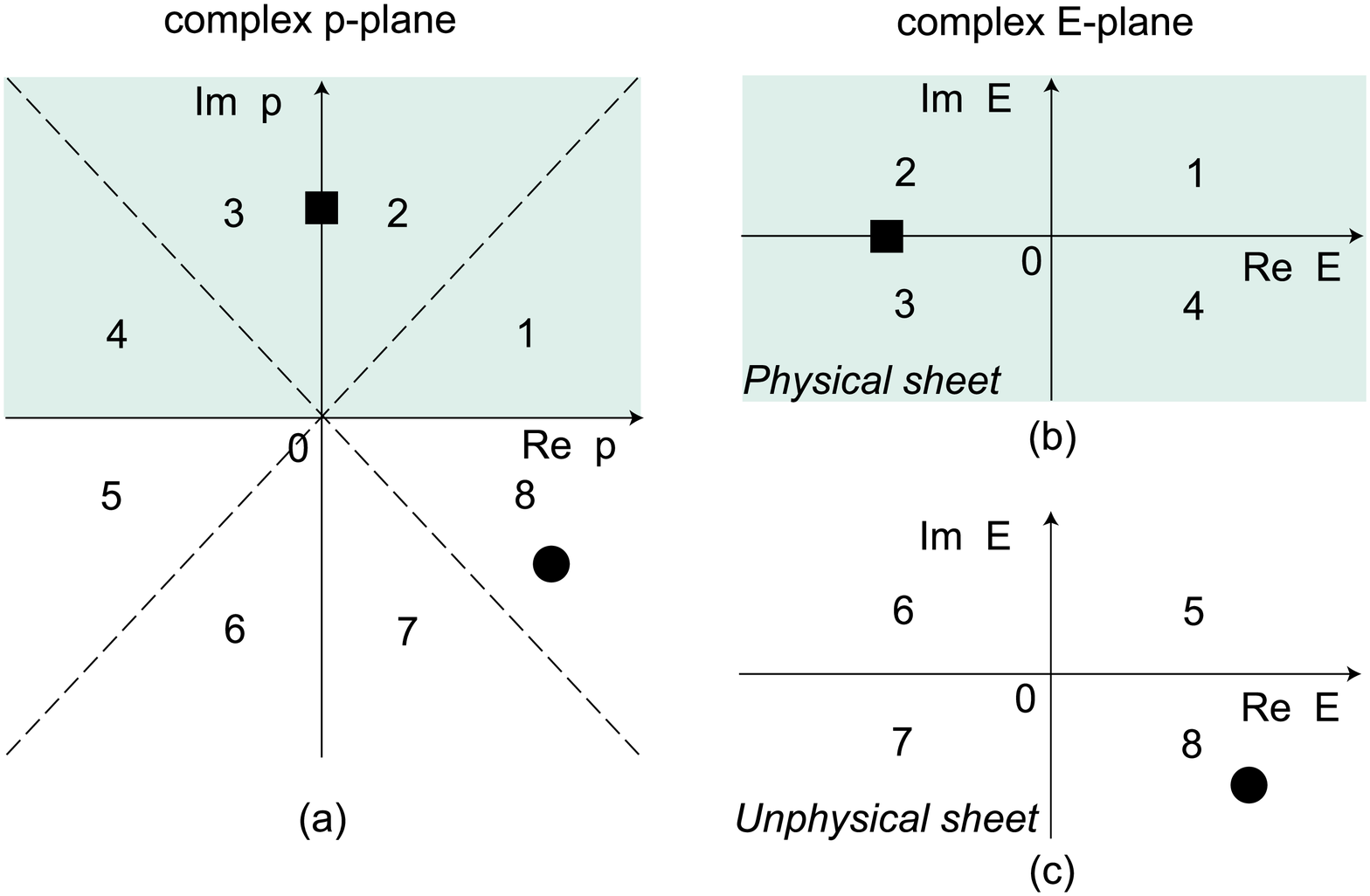}
\caption{The complex momentum $p$-plane ((a)) and its corresponding
complex energy $E$-plane which has a physical sheet ((b)) and
unphysical sheet ((c)). Their correspondence is indicated by
the same number. Solid squares (circles) represent the  bound state (resonance) poles.}
\label{fig:ep-sheet}
\end{center}
\end{figure}

It is useful to first  briefly recall how the resonances are defined
in textbooks. By analytic continuation, the scattering T-matrix can be 
defined on the complex energy plane. 
Its analytical structure is well studied\cite{newton1,newton,peierls59,
couteur60, kato65}
 for the non-relativistic two-body scattering.
For the single-channel case, the scattering
T-matrix is a  single-valued function of momentum $p$
on the complex  momentum p-plane, but is
a double-valued function of energy $E$ on the 
complex  energy E-plane because
of the quadratic relation $p=|2mE|^{1/2}e^{i\phi_{E}/2}$.
Therefore the complex E-plane is composed of two Riemann sheets.
The physical ($p$) sheet is defined by specifying the range of
phase $0 \le \phi_E \le  2\pi $,
and the un-physical ($u$) sheet
by $2\pi \le \phi_E \le 4\pi$.  As illustrated in Fig. \ref{fig:ep-sheet},
the  shaded area with $\mathrm{Im}\, p > 0$ of the  upper part of Fig. 1(a)
corresponds to the physical $E$-sheet shown in Fig. 1(b).
Similarly, the unphysical $E$-sheet shown in Fig. 1(c)
corresponds to the
$\mathrm{Im}\, p <0$ area in the lower part of Fig. 1(a).
On the physical sheet, the only possible singularities  are on the
real E-axis :
the bound state poles (solid square in Fig. 1(b)) below the threshold 
energy $E_{th}$
and the unitarity cut from $E_{th}$ to infinity.
On the unphysical sheet, a pole (solid circle in Fig. 1(c)) 
on the lower half plane and 
$\mathrm{Re}\, E_{pole} > E_{th}$
corresponds to a
resonance.  From unitarity and analyticity of S-matrix, each resonance pole
has an accompanied pole, called conjugate pole, which exists on the upper half
of the un-physical sheet.
A resonance pole is due  to the mechanism
: an unstable system is formed and decay subsequently during the collision.
The mathematical details of this interpretation
 can be found in textbooks, such as in chapter
8 of Goldberger and Watson\cite{gw}.

For multi-channel case, the analytic structure of the scattering
T-matrix  becomes more complex
\cite{newton,peierls59,couteur60,kato65}. 
We postpone the discussion on this
until section III where 
a two-channel Breit-Wigner form of scattering amplitude will be used
to give a pedagogical explanation.

The essential point of the Time-Delay  and Speed-Plot methods is that
the resonance poles discussed above can be determined from the partial-wave
amplitudes defined on the physical real energies.
The concept of time delay was originally introduced 
by Eisenbud\cite{eisenbud} and discussed by Wigner\cite{wigner55}, 
Dalitz and Moorehouse\cite{dalitz70}, and Nussenzveig\cite{nussen72}.
It is defined as the difference between the time in which a wave packet
passes through an interaction region and the time spent by a free wave packet passing though
the same distance.
It was generalized to the time delay matrix in multi-channel problems and
discussed further in terms of a lifetime matrix by Smith \cite{smith60}.
Kelkar et al.\cite{kelkar03b, kelkar04} applied this multi-channel 
formulation to develop a practical Time-Delay  method
to extract hadron resonances.

The Speed-Plot  method was developed by 
Hoehler\cite{hoehler92,hoehler93}
 to extract the nucleon resonances from the  $\pi N$ partial-wave amplitudes.
It is also based on the concept of time delay, but he identified resonances 
by "speed" which is the absolute value of 
the energy derivative of the scattering amplitude.
The speed  is
always positive by its definition while
the time delay  can be negative and becomes "time advance" \cite{kelkar03a}.

It is generally assumed that both  
the  Speed-Plot  and Time-Delay  methods 
give  approximate positions of the resonance poles which are rigorously defined
by the analytic continuation of  scattering amplitudes.
However their validity is not clear at all.
Moreover it is not clear which Riemann sheet the pole positions extracted from 
using these methods belong to.
In this paper we analyze several exactly solvable models to clarify these
questions.
In particular, we find that these two methods give poles
on different Riemann sheets. 

We then develop an analytic continuation method for extracting
the resonance parameters within a recently developed 
Hamiltonian formulation\cite{msl} of meson-nucleon reactions. 
We first establish
the method by using several exactly solvable models. The main task
 here is to handle the singularities associated with the unstable particle 
channels such as $\pi \Delta$, $\rho N$, and $\sigma N$. We then apply
 the method to extract the nucleon resonances in some 
partial waves of $\pi N$ scattering within 
the dynamical coupled-channels model 
developed in Ref.\cite{jlms}.  The analytical structure of the resonance
propagator is also analyzed within this model.

In section II, we describe the formula for applying
 the Speed-Plot (SP) and 
Time-Delay (TD) methods.  These
 two methods are then analyzed and tested in section III by
using the commonly used two-channel Breit-Wigner form 
of  S-matrix.
In section IV, we  develop an analytic continuation method 
by using several
exactly solvable  resonance
models and further
test the SP and TD methods. Section V is devoted to
resolve the complications due to couplings with 
unstable particle channels.
In section VI, we present
the results from applying the developed analytic continuation  method to 
extract the nucleon resonances in some partial waves
within the $\pi N$ model of Ref.\cite{jlms}.
A summary is given in section VII. 

\section{Formula for Time-Delay and Speed-Plot Methods}

The objective of the Time Delay (TD) and Speed-Plot (SP) methods is to
determine the mass $M_R$ and width $\Gamma_R$ of a resonance from the S-matrix
of reactions. 
In this section, we will not explain how these two methods
were introduced, as briefly discussed in section I. Rather we only give
their formula in practical applications.

For the single-channel elastic scattering case, the time delay of the 
outgoing wave packet with respect
to the non-interacting wave packet in a given partial wave
is defined\cite{wigner55,dalitz70,nussen72} as
\begin{eqnarray}
\Delta t (E) & = & \mathrm{Re}\left(-i\frac{1}{S(E)}\frac{dS(E)}{dE}\right)\,,
\label{eq:td-1}
\end{eqnarray}
where the S-matrix is related to
the phase  shift $\delta$ by 
\begin{eqnarray}
S(E)=e^{2i\delta}\,.
\label{eq:phaseshift}
\end{eqnarray}
Eqs. (\ref{eq:td-1}) and (\ref{eq:phaseshift})
lead to a simple expression of time delay
\begin{eqnarray}
 \Delta t =2\frac{d\delta}{dE}\,.
\label{eq:td-2}
\end{eqnarray}

The time-delay (TD) method is to find the resonance mass $M_R$ by 
finding  the maximum  
of the time delay
\begin{eqnarray}
\frac{d\Delta t}{dE}\mid _{E=M_R} =0.
\label{eq:td-e}
\end{eqnarray}
As an ideal example, we evaluate the time delay
for the S-matrix defined  by the well known Breit-Wigner resonance formula
\begin{eqnarray}
S(E) & = & \frac{E - M_R - i \Gamma_R/2}{E - M_R + i \Gamma_R/2}, \label{eq-bw}
\end{eqnarray}
Eq. (\ref{eq:td-1}) then leads to
\begin{eqnarray}
\Delta t & = & \frac{\Gamma_R}{(E-M_R)^2 + \Gamma_R^2/4},
\end{eqnarray}
which obviously takes a maximum at $E=M_R$ and hence $M_R$ is defined as the
resonance mass.  It also gives the following simple physical interpretation
of the width $\Gamma_R$ in terms of the time delay of the wave packet passing
through the interaction region
\begin{eqnarray}
\Delta t|_{E=M_R} & = & \frac{4}{\Gamma_R}.
\end{eqnarray}

Here and in the rest of this paper,
the normalization is chosen
such that the S-matrix is related to the T-matrix by
\begin{eqnarray}
S(E)=1+2iT(E) \,.
\label{eq:s-t}
\end{eqnarray}
We then note that for the S-matrix Eq. (\ref{eq-bw}), the width
can also be expressed in terms of T-matrix  as
\begin{eqnarray}
\frac{\Gamma_R}{2}=\left[\frac{|T|}{|\frac{dT}{dE}|}\right]_{E=M_R}.
\label{eq:sp-width1}
\end{eqnarray}

In the analysis of $\pi N$ scattering, Hoehler\cite{hoehler92,hoehler93}
 introduced a speed plot (SP) method. The speed is defined as
\begin{eqnarray}
\mathrm{sp}(E) & = & \left|\frac{dT}{dE}\right|.
\label{eq:sp-1}
\end{eqnarray}
Using Eqs. (\ref{eq:phaseshift})-(\ref{eq:td-2}), Eq. (\ref{eq:sp-1}) leads to 
\begin{eqnarray}
\mathrm{sp}(E) = \frac{|\Delta t|}{2}.
\label{eq:sp-td}
\end{eqnarray}
Thus the speed is also related to the time delay of the wave packet. 
The SP method defines the resonance mass $M_R$ by
the maximum of the speed
\begin{eqnarray}
\frac{d}{dE} \mathrm{sp}(E)\mid_{E=M_R} =0.
\label{eq:sp-e}
\end{eqnarray}
From Eqs. (\ref{eq:td-e}), (\ref{eq:sp-td}), and (\ref{eq:sp-e}),
it is obvious that the speed plot and time delay will give 
the same resonance mass for the single-channel elastic scattering case.

For the multi-channel reactions with $N$ open channels, the S-matrix
becomes a $N \times N$ matrix $\hat{S}(E)$. 
Smith\cite{smith60} introduced a life-time matrix $\hat{Q}$ 
defined by
\begin{eqnarray}
\hat{Q}(E) & = & -i \hat{S}(E)^\dagger \frac{d\hat{S}(E)}{dE}.
\label{eq:td-smith}
\end{eqnarray}
The above equation can be considered as
an extension of Eq. (\ref{eq:td-1}) of the single-channel case.
The trace of the life-time matrix can be 
expressed in terms of eigen phases $\delta_i$ of the S-matrix\cite{hazi,iga04,haber}
\begin{eqnarray}
\mathrm{Tr}\, \hat{Q}(E) & = & 2 \frac{d(\sum_i \delta_i)}{dE}.
\label{eq:td-smith-1}
\end{eqnarray}
The resonance mass $M_R$ is then obtained by finding
\begin{eqnarray}
\frac{d\mathrm{Tr}\, \hat{Q}(E)}{dE}\mid_{E=M_R} =0.
\label{eq:td-smith-2}
\end{eqnarray}
However,  the eigen phases $\delta_i$ can be obtained only when 
we know all of the
S-matrix elements associated with all open channels. In practice,
only the elastic scattering amplitude and a few
of the inelastic amplitudes can be extracted from the data.
Therefore Eqs. (\ref{eq:td-smith-1})-(\ref{eq:td-smith-2}) of Smith's TD method
can not be used rigorously in practice.

Here we focus on the TD method  
used by Kelkar et al.\cite{kelkar03a,kelkar03b,kelkar04}.  They defined
the time delay 
 for the  channel $i$ 
only by the diagonal component $S_{ii}$ of the S-matrix and its derivative
\begin{eqnarray}
\mathrm{td}(E)& = & \mathrm{Re}\left(-i\frac{1}{S_{ii}}\frac{d S_{ii}}{dE}\right). \label{td-mul}
\label{eq:td-kel-1}
\end{eqnarray}
Obviously, 
 this method is  identical to Eq. (\ref{eq:td-1}) of
the single channel case except that the S-matrix element here is
$S_{ii}= \eta e^{2i\delta}$ with $\eta$ denoting the inelasticity.
Eq. (\ref{eq:td-kel-1}) can be considered as an approximation of
Eq. (\ref{eq:td-smith}) by neglecting the inelastic channels in summing the
intermediate states.
In Refs. \cite{kelkar03a,kelkar03b,kelkar04}, the resonance mass $M_R$ 
 is determined by the maximum of the time delay
\begin{eqnarray}
\frac{d}{dE} \mathrm{td}(E)\mid_{E=M_R}&=&0 
\label{eq:td-kel-2a}, \\
\end{eqnarray}
and the width $\Gamma_R$ by
\begin{eqnarray}
\mathrm{td}(M_R \pm \Gamma_R/2)& = & \mathrm{td}(M_R)/2.
\label{eq:td-kel-2b}
\end{eqnarray}

The SP method for multi-channel case is simply to define the speed by
 the diagonal
$T_{ii}$ of the T-matrix
 \begin{eqnarray}
\mathrm{sp}(E) & = & \left|\frac{d T_{ii}}{dE}\right|,
\label{eq:sp-nch}
\end{eqnarray}
and define\cite{hoehler92} the width $\Gamma_R$ by assuming that 
the T-matrix element can be parametrized as
\begin{eqnarray}
T_{ii}(E) & = & T_b(E) + z(E)\frac{\Gamma_R/2}{E - M_R + i\Gamma_R/2}.
\label{eq:sp-t}
\end{eqnarray}
Here $T_b$ is an non-resonant amplitude and the resonant amplitude
is defined by the resonance mass $M_R$,  the width $\Gamma_R$, and
a complex residue $z(W)$. 
By assuming that  the energy dependence of 
$T_b(W)$, $z(W)$, and $\Gamma_R$  can be neglected
at energies near $M_R$, 
Eqs. (\ref{eq:sp-nch}) and (\ref{eq:sp-t}) obviously
 satisfies Eq. (\ref{eq:sp-e}) and lead to the following condition
\begin{eqnarray}
\mathrm{sp}(M_R \pm \Gamma_R/2) & =& \mathrm{sp}(M_R)/2.  \label{speed-width}
\label{eq:sp-width}
\end{eqnarray}
Eqs. (\ref{eq:sp-nch}), (\ref{eq:sp-e}) and (\ref{speed-width}) 
are used in applying the SP method to extract the resonance
mass $M_R$ and width $\Gamma_R$ from the partial-wave amplitudes.  
Eq. (\ref{eq:sp-t})
is not needed in practice, but is an essential assumption of SP method.

%

\section{Analysis of Speed-Plot and Time-delay Methods}

To examine the SP and TD methods, we consider a commonly used two-channel
Breit-Wigner (BW) amplitude which can be derived\cite{newton,kato65,fujii,fujii75}
 from the analytical property of the S-matrix. To make the contact with what
we will discuss in the rest of this paper, we will indicate here how this 
amplitude can be derived from a Hamiltonian formulation of 
meson-baryon  reactions, such as that developed 
in Ref.\cite{msl}. 

It is sufficient to consider the simplest two-channel case 
with a non-relativistic two-particle
Hamiltonian defined by
\begin{eqnarray}
H & = & H_0 + V.
\end{eqnarray}
In the center of mass frame $H_0$ can be written as
\begin{eqnarray}
H_0 = \sum_{i} \ket{i}[m_{i1} + m_{i2} + \frac{p^2}{2\mu_i}]\bra{i},
\end{eqnarray}
where $m_{ik}$ is the mass of $k$-th particle in channel $i$, and
$\mu_i =m_{i1}m_{i2}/(m_{i1}+m_{i2})$ is the reduced mass.
In each partial-wave, the S-matrix is a $2\times 2$ matrix and can be
written  
\begin{eqnarray}
S = \frac{1 - i \pi \rho K}{1 + i \pi \rho K},
\label{eq:s-kmx}
\end{eqnarray}
where $\rho$ is the density of state,  and
 the K-matrix, which is also a $2\times 2$ matrix,
 is defined  by the following Lippmann-Schwinger equation
\begin{eqnarray}
K(E) = V + V \frac{P}{E - H_0}K(E).
\end{eqnarray}
Here $P$ means taking the principal-value of the integration over the
propagator.

We now consider the on-shell matrix element of the S-matrix Eq. 
(\ref{eq:s-kmx}). 
If the on-shell momentum is denoted as $p_i$ for channel $i$, we
then have 
\begin{eqnarray}
\braket{i|1 \pm i \pi \rho K |j} & = &
  \delta_{i,j} \pm i \pi \rho_i K_{i,j},
\end{eqnarray}
where $\rho_i = p_i\mu_i$.
The  S-matrix element of the $1\rightarrow 1$
elastic scattering is then of the following explicit form
\begin{eqnarray}
S_{11} & = & \frac
{(1 - i\pi\rho_1 K_{11})(1 + i\pi\rho_2K_{22})-\pi^2\rho_1\rho_2K_{12}K_{21}}
{(1 + i\pi\rho_1 K_{11})(1 + i\pi\rho_2K_{22})-\pi^2\rho_1\rho_2K_{11}K_{22}}.
\label{eq:s-matrix-11}
\end{eqnarray}
If we
assume that at energies near the resonance energy the
K-matrix can be approximated as
\begin{eqnarray}
K_{ij} \sim \frac{g_ig_j}{E - M},
\end{eqnarray}
where $M $ is a mass parameter which is a real number,
Eq. (\ref{eq:s-matrix-11}) can then be written as
\begin{eqnarray}
S_{11} & = & \frac{E - M - i p_1 \gamma_1 + i p_2 \gamma_2}
                  {E - M + i p_1 \gamma_1 + i p_2 \gamma_2}.
\label{eq:bw-s}
\end{eqnarray}
Here we have defined $\gamma_i = \pi g_i^2 \mu_i >0$.
If we further assume that 
$\gamma_i$ is independent of scattering
energy, Eq. (\ref{eq:bw-s}) is the commonly
used two-channel Breit-Wigner formula\cite{fujii,fujii75,badalyan82}.
In the rest of this section, we will follow these earlier works
and treat
$\gamma_1$ and $\gamma_2$ as energy independent parameters of the
model.

Since the scattering T-matrix is related to the S-matrix by
\begin{eqnarray}
S_{11}(E)= 1 +2i T_{11}(E),
\end{eqnarray}
Eq.(\ref{eq:bw-s}) leads to 
\begin{eqnarray}
 T(E)&=& T_{11}(E) \nonumber \\
&=& \frac{-\gamma_1 p_1}{E-M+i\gamma_1 p_1 +i \gamma_2 p_2}.
\label{eq:bw-t}
\end{eqnarray}
From nowon we use the notation $T(E)$ for the $1\rightarrow 1$ 
amplitude $T_{11}(E)$.

\subsection{Analytic Properties of the S-matrix}
Within the two-channels Breit-Wigner model specified above, we will
analyze in this subsection 
the analytic properties of the S-matrix
on the complex energy E-plane. This will also allow
us to explain clearly some 
terminologies which are commonly seen  but  often not explicitly
explained in the literatures on resonance extractions.

The on-shell momenta $p_i$ for channel $i$ is defined by
\begin{eqnarray}
E_i=\frac{p^2_i}{2\mu_i} ,
\label{eq:p-on-1}
\end{eqnarray}
where
\begin{eqnarray}
E_i=E-( m_{i1}+m_{i2}).
\label{eq:p-on-2}
\end{eqnarray}
We can define the threshold variable $\Delta$ between two channels by
\begin{eqnarray}
E_1  = \frac{p_2^2}{2\mu_2} + \frac{\Delta^2}{2\mu_1},
\label{eq:p-on-3}
\end{eqnarray}
where 
\begin{eqnarray}
\frac{\Delta^2}{2\mu_1} & = & m_{21} + m_{22} - m_{11} - m_{12}
\label{eq:p-on-4}
\end{eqnarray}
is the threshold energy of the second channel.

The momenta at poles of the S-matrix Eq. (\ref{eq:bw-s})
 can be determined by solving
\begin{eqnarray}
E - M + i p_1 \gamma_1 + i p_2 \gamma_2 =0.
\label{eq:s-pole}
\end{eqnarray}
By using Eqs. (\ref{eq:p-on-1})-(\ref{eq:p-on-4}),
 the above equation can be written as
\begin{eqnarray}
p_1^4 + a p_1^3 + b p_1^2 + c p_1 + d = 0,
\label{eq:pole}
\end{eqnarray}
where
\begin{eqnarray}
a & = & i 4 \mu_1 \gamma_1, \\
b & = & 4\mu_1(\mu_2 \gamma_2^2 - M - \mu_1 \gamma_1^2),\\
c & = & -8i \mu_1^2 M\gamma_1,\\
d & = & 4(\mu_1^2 M^2 - \mu_1\mu_2 \gamma_2^2 \Delta^2).
\end{eqnarray}
Eq. (\ref{eq:pole}) means that  the Breit-Wigner
amplitude Eq. (\ref{eq:bw-s}) has four poles.  
Each pole is specified by two on-shell momenta
 $P_{\alpha}= (p_{1\alpha},p_{2\alpha})$ with $\alpha=1,2,3,4$.
 The analytic properties of the S-matrix Eq. (\ref{eq:bw-s})
depends on how these poles are located on the complex energy $E$-plane. 
As we explained in section I, the energy plane for each channel
has two Riemann sheets because of the quadratic relation Eq. (\ref{eq:p-on-1})
between the momentum
$p_i$ and energy $E_i$; namely
$p_i=\sqrt{2\mu_i|E_{i}|}e^{i\phi_i/2}$ for $i=1,2$. 
For each channel, the physical ($p$) sheet is defined by specifying the range of
phase $0 \le \phi_i \le  2\pi $, 
and the un-physical ($u$) sheet 
by $2\pi \le \phi_i \le 4\pi$. The correspondence between  the momentum
$p_i$-plane
and the energy $E_i$-plane is similar to that
illustrated in Fig. \ref{fig:ep-sheet}.
For the considered
two-channel case, we thus have
four energy sheets specified by the signs of $\mathrm{Im}\, p_1$
 and $\mathrm{Im}\, p_2$:
$pp$, $up$, $uu$, and $pu$, as shown in Fig.2. 
Thus each of four poles $P_{\alpha}= (p_{1\alpha},p_{2\alpha})$
  can be on one of these
 E-sheets.

\begin{figure}
\begin{center}
\includegraphics[width=12cm]{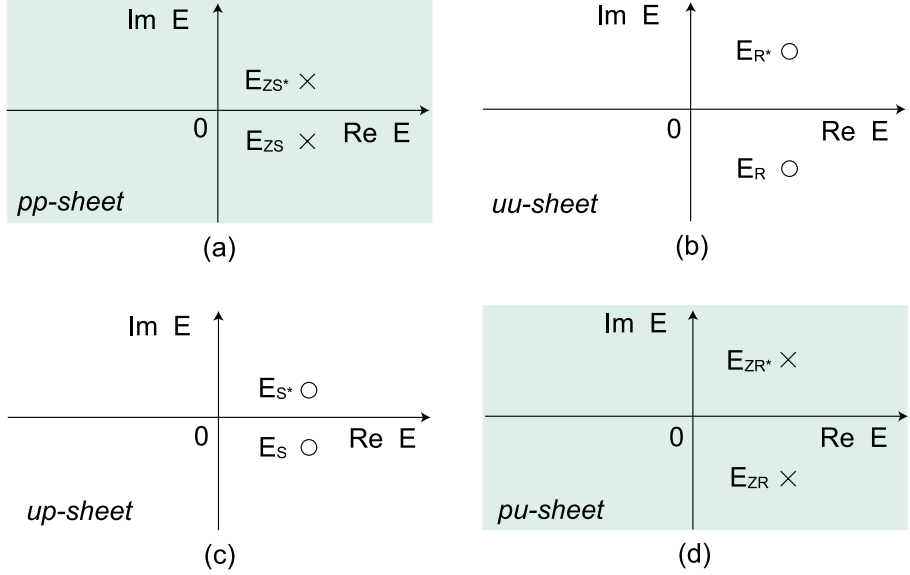}
\caption{Poles and zeros  of the simplified two-channel
Breit-Wigner form ($\mu_1=\mu_2$ and $\Delta=0$)
of S-matrix on the complex E-plane which has
$pp$ , $uu$ , $up$ and $pu$ sheets.
The open circles on the $uu$-sheet ((b)) are the
resonance pole $E_R$ and its conjugate pole $E_R^*$.
The open circles on the $up$-sheet ((c)) are the
shadow pole $E_S$ and its conjugate pole $E_S^*$.
The crossed on the $pp$-sheet ((a)) are the zero $E_{ZS}$ and its conjugate
$E^*_{ZS}$ which are at the same energies of
the shadow poles $E_{S}$ and $E^*_{S}$.
The crossed on the $pu$-sheet ((d))
are the zero $E_{ZR}$ and its conjugate
$E^*_{ZR}$ which are at the same energies of
the resonance poles $E_{R}$ and $E^*_{R}$.}
\label{pole-2}

\end{center}
\end{figure}

To be more specific, we now consider the case which is most relevant to
our study of nucleon resonances. That is the  
$\mathrm{Re}\, E_1 >0$ and $\mathrm{Re}\, E_2 >0 $ case that the poles are all 
above the thresholds of both channels.  
From Eq. (\ref{eq:s-pole}),
we immediately notice that if $(p_{1a},p_{2a})$ with $E=E_a$ is one of the
solutions,
 $(-p_{1a}^*,-p_{2a}^*)$ with $E=E^*_a$ is also a solution.
Therefore the four poles determined by Eq. (\ref{eq:pole}) can be grouped
into two pairs. In the following discussions,
they are denoted as  $(E_a, E^*_a)$ and $(E_b,E^*_b)$. 
Without  losing  generality,
one can assume that one of the poles
is in the range of ($\mathrm{Re}\, p_1>0, \mathrm{Re}\, p_2 >0$)
and the other in the range of ($\mathrm{Re}\, p_1<0, \mathrm{Re}\, p_2 <0$).
If the first pole $(p_{1a},p_{2a})$ is in 
the region where ($\mathrm{Re}\, p_{1a}>0$, $\mathrm{Re}\, p_{2a}>0$) and
 ($\mathrm{Im}\, p_{1a}<0$, $\mathrm{Im}\, p_{2a}<0$) , it 
is a pole, denoted as $E_R$,  on the $uu$-sheet of Fig. \ref{pole-2}.
This pole is usually called  the resonance pole and is  closer than other poles
on $up$ or $pu$-sheets  
to the physical $pp$-sheet, as  will be explained later. 
In the Hamiltonian formulation considered in this work and the
well developed collision theory, a resonance pole can be mathematically
derived\cite{gw} from  the mechanism 
that an unstable system is formed and decay subsequently during the collision.
 The resonance pole $E_R$ has an accompanied pole $E^*_R$
at $(-p_{1a}^*, -p_{2a}^*)$ which is also on the $uu$-sheet
as shown  in Fig. \ref{pole-2}. $E_R^*$ is called the 'conjugate pole' of $E_R$.

The second pole at $(p_{1b},p_{2b})$ with $\mathrm{Im}\, p_{1b}<0$
 and $\mathrm{Im}\, p_{2b}>0$ ( $\mathrm{Im}\, p_{1b}>0$ and $\mathrm{Im}\, p_{2b}<0$)
 may be on the $up$-sheet ($pu$-sheet), depending on the parameters 
$\gamma_1$ and $\gamma_2$.
This pole is called  the shadow pole\cite{eden64}.
A shadow pole on $up$-sheet and its conjugate pole are 
$E_S$ and $E^*_S$ in  Fig. \ref{pole-2}.

We now note that in this simple BW model, the zeros of 
the S-matrix Eq. (\ref{eq:bw-s}), where $S_{11}(E)=0$,
is defined by its numerator
\begin{eqnarray}
E - M - i p_1 \gamma_1 + i p_2 \gamma_2 = 0.  
\label{eq:s-zero-1}
\end{eqnarray}
The above equation can be cast into the form of Eq. (\ref{eq:s-pole}) by
 simply replacing $p_1$ by $-p_1$.
Thus solutions of Eq. (\ref{eq:s-zero-1}), called the zeros of S-matrix, can 
be readily 
obtained from the solutions 
$(p_{1a},p_{2a})$ and $(p_{1b},p_{2b})$
of Eq. (\ref{eq:s-pole}). They are 
$(-p_{1\alpha},p_{2\alpha})$ with $E_\alpha$ 
and  $(p_{1\alpha}^*,-p_{2\alpha}^*)$ with $E^*_\alpha$ for $\alpha=a,b$.
The zero at $(-p_{1b},p_{2b})$ is on the $pp$-sheet,
denoted as $E_{ZS}$ and $E^*_{ZS}$ in Fig. \ref{pole-2}.
Similarly, the zero at $(-p_{1a},p_{2a})$ is on the $pu$-sheet,  
shown as $E_{ZR}$ together with its conjugate  $E^*_{ZR}$ in
Fig. \ref{pole-2}. Note that Fig. \ref{pole-2} is for the case that
the parameters $\gamma_1$ and $\gamma_2$ are chosen such that the
shadow poles $E_{S}$ and its conjugate $E^* _{S}$ are on the $up$-sheet.
For other possible $\gamma_1$ and $\gamma_2$, the pole positions could be
different from what are shown in Fig. \ref{pole-2},
 but their close relations, as discussed above, are the same.

\begin{figure}
\begin{center}
\includegraphics[width=12cm]{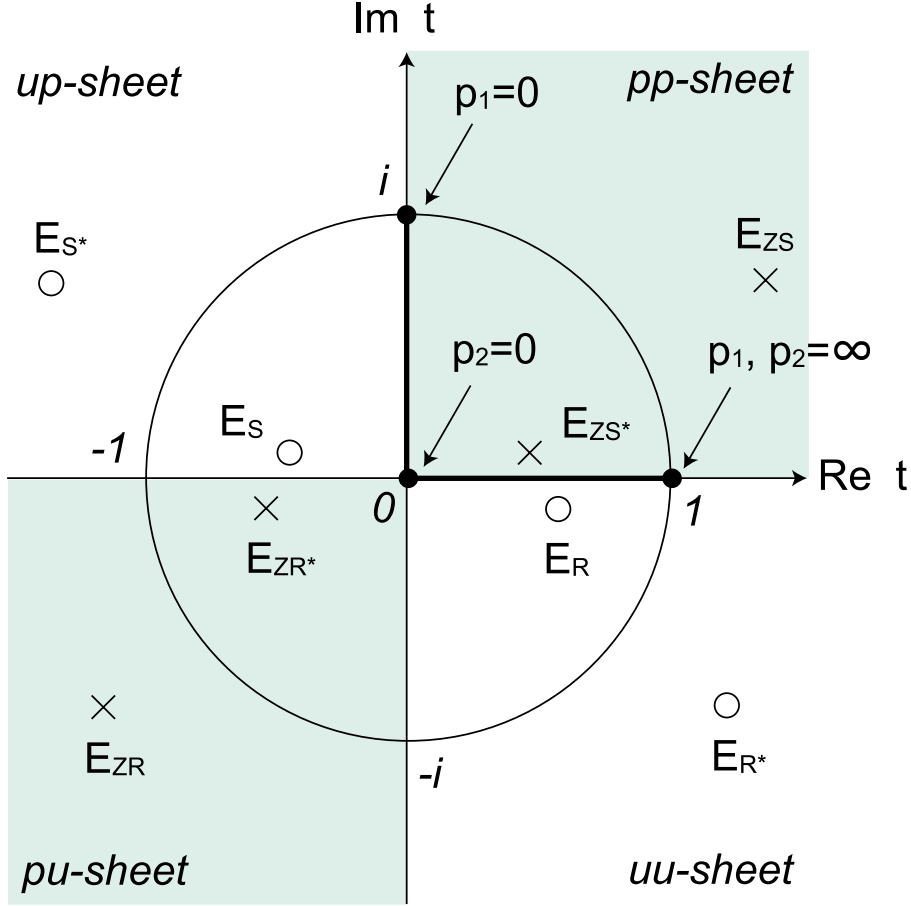}
\caption{The poles and zeros of S-matrix shown in Fig.2 are displayed
on the complex t-plane defined by Eqs. (\ref{eq:t2}) and (\ref{eq:t2b}).} 
\label{fig:t-plane}
\end{center}
\end{figure}

From the above analysis, it is clear that the poles and zeros of the S-matrix
are closely related. Their locations  on the 4 Riemann sheets can 
be conveniently
displayed on  one complex plane by introducing
a variable $t$\cite{newton-book}
\begin{eqnarray}
p_1 & = & \Delta \frac{1 + t^2}{1 - t^2}, \label{eq:t2}  \\ 
p_2 & = & 2\Delta\sqrt{\frac{\mu_2}{\mu_1}}\frac{t}{1 - t^2}.
\label{eq:t2b}
\end{eqnarray}
Hence each point  in the $t-$plane corresponds to a set of $(p_1,p_2)$.
In Fig. \ref{fig:t-plane}, the resonance position $E_{R}$ and shadow $E_{S}$ 
poles
and their conjugate poles and zeros of S-matrix 
$E_{ZS}, E_{ZR}, E_{ZS}^*, E_{ZR}^*$
are shown on t-plane. The physical S-matrix at real energies, which determine
 the observables,  are on the 
bold lines. The zero energy and the threshold of the second channel
correspond to  $t=i$ and $t=0$, respectively.
One can see that the resonance pole $E_{R}$ is closer 
than the shadow pole $E_{S}$
to the bold lines (S-matrix) and
hence can have the largest effect on the observables. 
Consequently, most of the rapid energy dependence of observables
are attributed to the resonance poles, not the shadow poles or the other poles
shown in Fig.\ref{fig:t-plane}. On the other hand, the zero $E^*_{ZS}$
of the S-matrix is also close to
the bold lines. As seen in the derivations given above,
this zero $E^*_{ZS}$ is closely related to shadow pole $E_{S}$. 
Thus the shadow poles
can also be related to the observables. 
Of course, which pole is most important in determining the
rapid energy dependence of observables also depends on the residues
of the T-matrix at the pole positions.

In the next subsection, we will use a further simplified BW form to
 explain more clearly the close relations between the poles and
zeros of the S-matrix. In particular we will
see explicitly that the
shadow pole  $E_S$ in Fig. \ref{pole-2}, which is not on the same Riemann sheet as $E_R$,
can be located by the zeros of the S-matrix.
More importantly, we will also see
  how the SP and TD methods work  both
analytically and numerically.

\subsection{Positions of poles}

For simplicity, we assume that the threshold energies of the two channels
 are the same and hence $\mu \equiv \mu_1=\mu_2$ and $\Delta=0$.
Therefore we have  simple relations  between energy and momenta :
$ E=p_1^2/2\mu=p_2^2/2\mu $ and $p \equiv p_1=\pm p_2$.
As discussed in the previous subsection, 
the four Riemann sheets  are classified by the sign of the
 imaginary part of the momentum; namely, 
 physical (unphysical) sheet is assigned by $\mathrm{Im}\, p >0$ $(\mathrm{Im}\, p<0)$. 
With the simplification $p \equiv p_1=\pm p_2$, 
we obviously have $p_1=p_2=p$ on the $pp$ and $uu$-sheets,
and $p_1=-p_2=p$ on $up$ and $pu$-sheets.

Let us start with the case of  $p_1=p_2=p$ for the  $pp$-sheet or
$uu$-sheet. The S-matrix element Eq. (\ref{eq:bw-s})
of the first channel 
can be written as
\begin{eqnarray}
 S_{11}(E) & = & \frac{E - M - i (\gamma_1 - \gamma_2)p}
                      {E - M + i (\gamma_1 + \gamma_2)p}.
\label{eq:s11-uupp}
\end{eqnarray}
It can be cast into the following more transparent form
\begin{eqnarray}
 S_{11}(E) & = & \frac{(p + p_S)(p - p_S^*)}{(p - p_R)(p + p_R^*)},
  \label{eq:s11-p}
\end{eqnarray}
with
\begin{eqnarray}
 p_R & = & \sqrt{2\mu M - (\mu\gamma_+)^2} - i \mu \gamma_+,
\label{eq:p-r} \\
 p_S & = & \sqrt{2\mu M - (\mu\gamma_-)^2} - i \mu \gamma_-,
\label{eq:p-s} 
\end{eqnarray}
where
\begin{eqnarray}
 \gamma_{\pm}  =  \gamma_1 \pm \gamma_2.
\end{eqnarray}
Remembering that we consider $\gamma_1 >0$ and $\gamma_2 >0$.
To make use of Fig.\ref{pole-2} in the following discussion, 
we consider the case that  $\gamma_1 > \gamma_2$ and hence
$\gamma_\pm >0$ and both $p_R$ and $p_S$ defined in 
Eqs. (\ref{eq:p-r})-(\ref{eq:p-s}) are associated with the unphysical $u$-sheet.
For the case of $\gamma_- <0$, $p_S$ is associated with the physical $p$-sheet
and the following presentation can be easily modified to account for this
case.

Clearly, Eq. (\ref{eq:s11-p}) means that
the S-matrix has a pole at $p_1=p_2=p=p_R$ on the $uu$-sheet with a resonance
energy
\begin{eqnarray}
E_R & = & \frac{p_R^2}{2\mu} = M - \mu\gamma_+^2 - 
                             i \gamma_+\sqrt{\mu(2M - \mu\gamma_+^2)}\,.
 \label{uu-pole}
\end{eqnarray}
Its conjugate pole $E^*_R$ is at $p_1=p_2 =p= -p_R^*$. 
The positions of $E_R$ and $E^*_R$ are shown in 
the upper right side of Fig.\ref{pole-2}.
Eq. (\ref{eq:s11-p}) also indicates that the zero of S-matrix
is at $p_1=p_2=-p_S$ which is on the $pp$-sheet because of 
$\mathrm{Im} (-p_S) > 0$. The energy of this zero of S-matrix is
\begin{eqnarray}
E_{ZS} & = & \frac{p_S^2}{2\mu} = M - \mu\gamma_-^2 - 
                             i \gamma_-\sqrt{\mu(2M - \mu\gamma_-^2)}.
\label{eq:e-zs}
\end{eqnarray}
Its conjugate  $E_{ZS}^*$ is at $p_1=p_2 = p_S^*$. The positions
of $E_{ZS}$ and $E_{ZS}^*$ are on the
$pp$-sheet, as shown in the upper left side of Fig.\ref{pole-2}.

We next consider the $p_1=-p_2=p$ case that the poles and zeros of
the S-matrix are on the 
$up$ or $pu$-sheets.
 The S-matrix Eq. (\ref{eq:bw-s}) for this case then takes the following form
\begin{eqnarray}
 S_{11}(E) & = & \frac{E - M - i (\gamma_1 + \gamma_2)p}
                      {E - M + i (\gamma_1 - \gamma_2)p}.
\label{eq:s11-uppu}
\end{eqnarray}
By comparing Eq. (\ref{eq:s11-uupp}) and  Eq. (\ref{eq:s11-uppu}) and using
the variables $p_R$ and $p_S$ defined by Eqs. (\ref{eq:p-r}) and (\ref{eq:p-s}),
Eq. (\ref{eq:s11-uppu}) 
can be written as
\begin{eqnarray}
 S_{11}(E) & = & \frac{(p + p_R)(p - p_R^*)}{(p - p_S)(p + p_S^*)}.
\label{eq:s11-up}
\end{eqnarray}
The above equation indicates that 
for the considered $\gamma_->0$, the S-matrix has a shadow pole at
$p_1=-p_2=p=p_S$ on the $up$-sheet. Thus its position
$E_{S}=p^2_S/(2\mu)$ is  identical to
$E_{ZS}$ of the zero of the S-matrix on the $pp$-sheet; namely
\begin{eqnarray}
E_S &=& E_{ZS} \nonumber \\
& = & \frac{p_S^2}{2\mu} = M - \mu\gamma_-^2 -
                             i \gamma_-\sqrt{\mu(2M - \mu\gamma_-^2)}.
\label{eq:e-s}
\end{eqnarray}
This means that the shadow pole $E_S$ on the $up$-sheet
can be found from searching for 
the zero $E_{ZS}$ of S-matrix on the $pp$-sheet.

Eq. (\ref{eq:s11-up}) also gives a zero of S-matrix at 
$p_1=-p_2=p=-p_R $ on $pu$-plane because $\mathrm{Im} (-p_R) >0$. 
Its energy  $E_{ZR}$ is also identical to $E_R$ defined above
\begin{eqnarray}
E_{ZR}&=& E_R \nonumber \\
&=&\frac{p_R^2}{2\mu} = M - \mu\gamma_+^2 -
                             i \gamma_+\sqrt{\mu(2M - \mu\gamma_+^2)}\,.
\label{eq:e-zr}
\end{eqnarray}
The positions of $E_S$ and
$E_{ZR}$ and their conjugates $E^*_S$ and $E^*_{ZR}$ are also in the
lower parts of
Fig.\ref{pole-2}.

From the above analysis  for the $\gamma_- >0$ case, 
we see that  the energies of the resonance poles
may be obtained by studying the poles of the S-matrix on the $uu$-sheet
and those of the shadow poles may be obtained from
the zeros of the S-matrix on the $pp$-sheet. 
The analysis for the $\gamma_- <0$ case  is similar. Here we only
mention that when $\gamma_- >0$ is changed to $\gamma_- <0$,
the shadow poles $E_S$ and $E^*_S$ on the $up$-sheet move to the $pu$-sheet 
and zeros $E_{ZS}$ and $E^*_{ZS}$ will be on the $uu$-sheet.

Now let us examine how the SP and TD methods can
find the poles defined  by the above exact expressions of the
two-channel BW amplitude. We first recall that
in applying the SP and TD methods, the energy $E$ and momentum $p_i$
in the S-matrix are restricted on the positive real-axis.
For the considered simplified case, we thus have $p_1=p_2=p$ and
the $T$-matrix Eq.(\ref{eq:bw-t}) 
then becomes
\begin{equation}
  T(E) = \frac{-\gamma_1 p}{E-M+i\gamma_+ p}. 
\label{t-pp}
\end{equation}
Our task is to examine whether
the resonance mass $M_R$ and width $\Gamma_R$ found by applying the SP and TD 
methods on the expression Eq.(\ref{t-pp})
are close to the real and imaginary parts of the
poles defined in the previous subsection.

According to Eqs. (\ref{eq:sp-e}) and (\ref{eq:sp-nch}),
the SP method finds the resonance mass $M_R$ by finding
 the maximum of the speed through the use of the condition
\begin{eqnarray}
 \left[\frac{d}{dE}\left | \frac{dT}{dE} \right |\right]_{E=M_R} = 0\,.
\label{eq:sp-sbw}
\end{eqnarray}
With the T-matrix Eq.(\ref{t-pp}), Eq. (\ref{eq:sp-sbw})
leads to the following equation
\begin{equation}
  3M_R^3+(3M+2\mu\gamma_+^2)M_R^2-(7M^2-6\mu M\gamma_+^2)M_R+M^3 = 0.
\end{equation}
This equation can be written in the dimension-less form as
\begin{equation}
  3\tilde{E}^3+(3+2\alpha)\tilde{E}^2-(7-6\alpha)\tilde{E}+1 = 0,
\label{eq:bw-pole-exp}
\end{equation}
with $\tilde{E}=M_R/M, \alpha=\mu\gamma_+^2/M_R$. With some inspections,
one can see that Eq. (\ref{eq:bw-pole-exp}) has 
real and positive $M_R $ solutions
 only in the
$ \alpha < 0.417 $ region.
Two of the three solutions are the maximum and minimum points of the speed, 
and the third one is less than 0.
The SP method defines the maximum point of the speed
as the resonance mass. We find that this solution can be expanded as
\begin{equation}
  [M_R]_{SP} = M - \mu\gamma_+^2 - \frac{1}{4M}(\mu\gamma_+^2)^2
  + O\left(\frac{(\mu\gamma_+^2)^3}{M^2}\right).
\end{equation}
Clearly, $[M_R]_{SP}$ equals to the real part of
 Eq. (\ref{uu-pole})
 if we neglect the second and higher order terms in 
the expansion in powers of
$ \mu\gamma_+^2/M $.
Therefore the SP method is accurate only under the condition that 
 $ \mu\gamma_+^2/M \ll 1 $.
Moreover, it is clear from the above equation that speed has no stationary
point for  $ \mu\gamma_+^2/M > 0.417 $ and therefore the SP method
will fail to find the pole even if there is a pole within the model.

We next turn to discussing the width $\Gamma_R$ obtained by the SP method.
It is evaluated by using 
Eq. (\ref{eq:sp-width1}). We find that it can also be expanded as 
\begin{equation}
\frac{[\Gamma_R]_{SP}}{2} \equiv
\frac{|T|_{[M_R]_{SP}}}{\left | \frac{dT}{dE} \right |_{[M_R]_{SP}} }
= \sqrt{\mu\gamma_+^2(2M-\mu\gamma_+^2)}
\left [ 1 - \frac{\mu\gamma_+^2}{2M}
      + O\left(\left(\frac{\mu\gamma_+^2}{M}\right)^2 \right)  \right ].
\end{equation}
Here again, if we neglect higher order terms of $\mu\gamma_+^2/M$, the
SP method 
can give the imaginary part of $E_R$ of the exact expression Eq. (\ref{uu-pole}).
As we have seen in Fig.\ref{pole-2}, the two-channel BW S-matrix   
has two pairs of poles on the $uu$-sheet and the $up$-sheet.
However the SP method can only find the pole $E_R$ on the $uu$-sheet.

\begin{figure}
\begin{center}
\includegraphics[width=7.5cm]{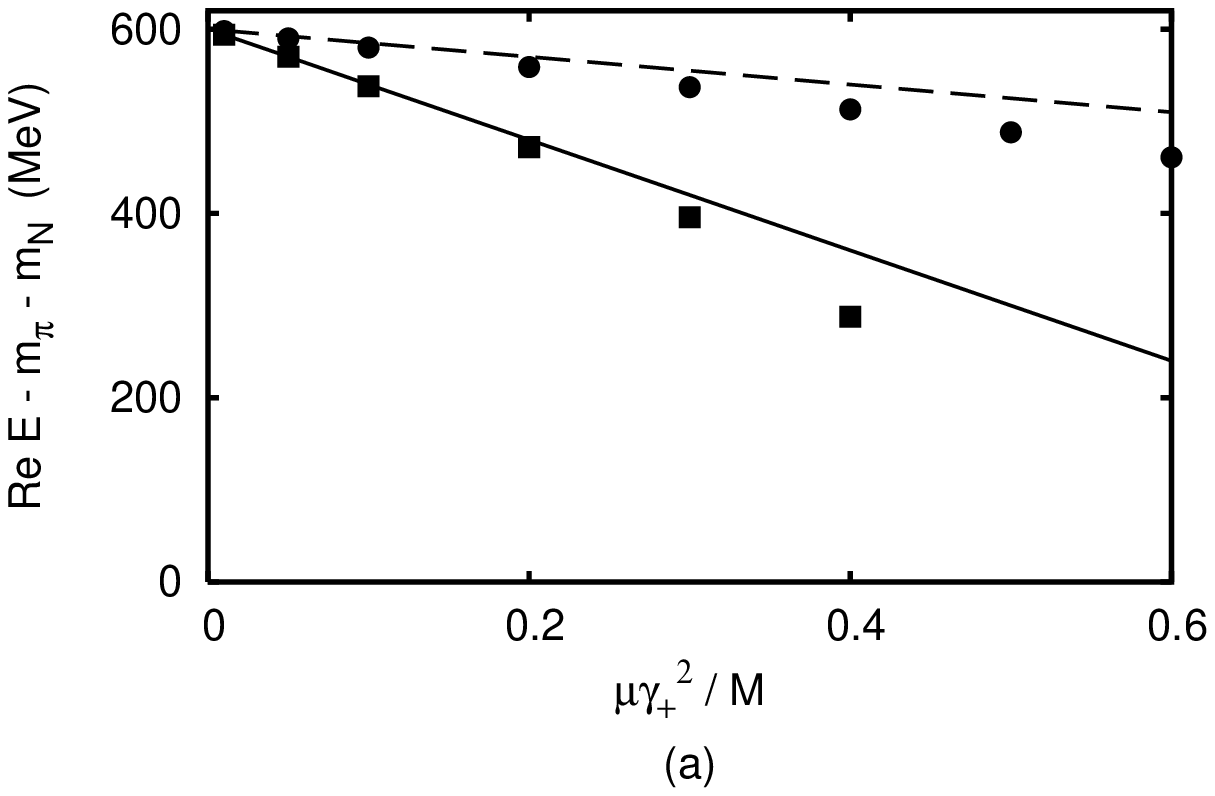}
\includegraphics[width=7.5cm]{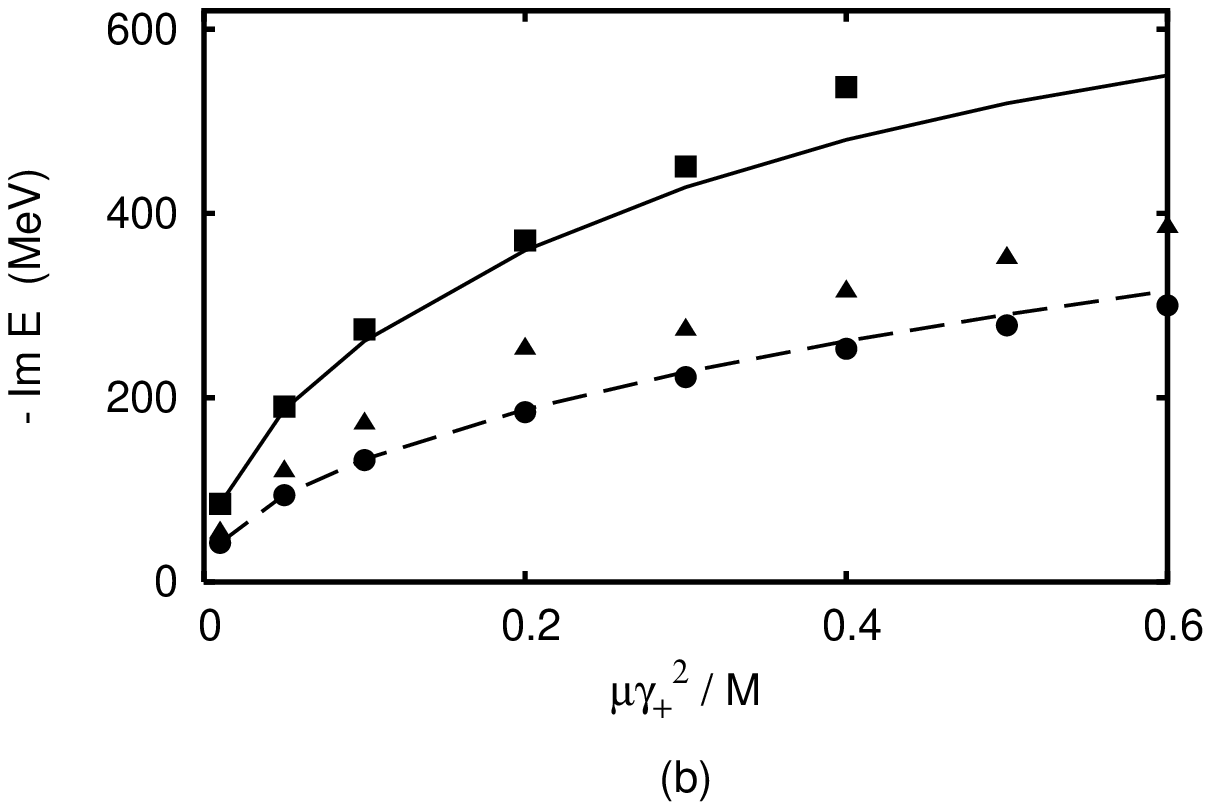}
\caption{
The $\mu\gamma_+^2/M$ dependence of the pole positions on the
$uu$-sheet (solid curve) and $up$-sheet (dashed curve) of the
simplified two-channels Breit-Wigner form ($\mu_1=\mu_2$, $\Delta=0$)
 calculated at $\gamma_-/\gamma_+=0.5$. The real (imaginary) parts of poles
are shown in  (a) ( (b)) of the figure.
The solid squares ( solid circles) are the results obtained from
using the SP (TD)  method.
Triangles in (b) are
 the widths of the poles obtained using Eq. (\ref{eq:td-kel-2b}).
}
\label{bw-pole1}
\end{center}
\end{figure}

From the above analysis, it is clear that
the accuracy of SP method is  controlled 
by $\mu\gamma_+^2/M$. We examine this by using an example with
  $\mu  = \frac{m_{N}m_{\pi}}{m_{N}+m_{\pi}}$,
  $m_{N} = 938.5 \ \mbox{MeV}$ and
  $m_{\pi} = 139.6 \ \mbox{MeV}$, and
   $  M = m_N + m_\pi + 600 \ \mbox{MeV}$.
 In Fig.\ref{bw-pole1} 
 the solid curves are the
 pole positions on the $uu$-sheet.
They are obtained from evaluating 
the exact analytical formula Eq. (\ref{uu-pole})
for $\gamma_-/\gamma_+=0.5$ and varying $\mu\gamma_+^2/M$ from $0.01$ to $0.6$.
With the same parameters we then apply the SP method to 
search for pole $(M_R, -i\Gamma_R/2)$ from the amplitude Eq. (\ref{t-pp}) 
numerically by using 
Eqs. (\ref{eq:sp-width1}), (\ref{eq:sp-1}) and  (\ref{eq:sp-e}).
The obtained pole positions
were the filled squares shown in Fig.\ref{bw-pole1}.
As expected, the SP method works very well for small $\mu\gamma_+^2/M$. However
the pole position from SP starts to deviate from the exact results
(solid curves) as $\mu\gamma_+^2/M$ increases.
There is no filled squares in Fig.\ref{bw-pole1} in the 
region   $\mu\gamma_+^2/M > 0.417$ because  the SP can not find a pole in
this region.
This is not because of the numerical accuracy of our calculation,
but is the intrinsic limitation of the SP method as, discussed above.

We now turn to investigating the TD method. We apply
Eq. (\ref{eq:td-kel-2a})
to search for the resonance mass $M_R$ from the amplitude
Eq. (\ref{t-pp}). Since it is not clear how to
interpret Kelkar's prescription Eq. (\ref{eq:td-kel-2b}),
 we assume that the S-matrix element $S_{11}$ is of the following form
\begin{eqnarray}
S_{11}(E) &=& \frac{E- M - i \Gamma_{S}/2}{E- M + i\Gamma_{R}/2} \\
&=& \eta e^{2i\delta}.
\end{eqnarray}
Then Eq. (\ref{eq:td-kel-2b}) leads to an improved expression for
the width $\Gamma_S$
\begin{eqnarray}
\frac{[\Gamma_R]_{TD}}{2} = \frac{\Gamma_S}{2}& \equiv &
 \frac{1\pm\eta }{2}\left |\frac{1}{\frac{d\delta}{dE}} \right |_{E=[M_R]_{TD}} ,
\label{eq:td-kel-2c}
\end{eqnarray}
where $+$ $(-)$ of $\pm$ is 
for the maximum(minimum) of the TD suggesting
 the  pole on $up$- ($pu$-) sheet.
We have found that the TD method, defined by Eq. (\ref{eq:td-kel-2a})
and Eq. (\ref{eq:td-kel-2c}) can only find the shadow poles
on the $up$-sheet which are given by
exact expression Eq. (\ref{eq:e-s}). The results from using the same
parameters specified above are also
 shown in Fig.\ref{bw-pole1}.
The dashed curves are from the exact expression
Eq. (\ref{eq:e-s}) and the solid dots are from applying
Eq. (\ref{eq:td-kel-2a}) and Eq. (\ref{eq:td-kel-2c}) to search numerically
for the poles from Eq. (\ref{eq:s11-uppu}).
Clearly TD method works very well in finding the shadow poles on $up$-sheet.
In the same figure the triangles are from using Kelkar's prescription
 Eq. (\ref{eq:td-kel-2b}). Obviously,
 our 
formula Eq. (\ref{eq:td-kel-2c}) works better.
We have also examined TD for $\gamma_-< 0$. In this case
the TD becomes negative and has minimum.
We apply Eq. (\ref{eq:td-kel-2a}) for the minimum of TD 
and find that TD works also well in finding the  pole on $pu$-sheet.

\section{Analytic continuation of resonance models}

With the analysis presented in the previous section, it is clear that
the empirical partial-wave amplitudes determined from experimental
data can not be blindly used to extract resonance parameters by using 
SP or TD methods. To make progress, one needs to construct a reaction
model to fit the data and then extract the resonance parameters by analytic
continuation within the model.
In this paper, we focus on a dynamical model\cite{msl} 
 which accounts for the main features of  meson production reactions in the 
nucleon resonance region. Our task in this section is to develop numerical
methods for finding the resonance poles from such models which do not have
analytical forms of their solutions. We will first consider the simplest
one-channel and one-resonance case, then two-channels and one-resonance, and 
finally two-channels and two-resonances cases. 
All of these models are exactly solvable such that their poles are known 
analytically and the developed numerical methods can be tested.
\subsection{One-channel, one-resonance}
To be specific, we consider 
the two-particle reactions defined by the following well known
isobar Hamiltonian 
in the center of mass frame 
\begin{eqnarray}
H = H_0 + H',
\end{eqnarray}
with
\begin{eqnarray}
H_0 &=&  [E_{1}(\vec{p}) +E_{2}(-\vec{p})] + \ket{N_0} M_0 \bra{N_0} \,,\\
H' &=&   \ket{g} \bra{g}\,,
\end{eqnarray}
where $M_0$ is the mass parameter
of a bare particle $N_0$ which can decay into two particle states through
the vertex interaction $g$ in $H'$, and
$E_{i}(p)=[m_{i}^2+p^2]^{1/2}$ is the energy of the $i$-th particle.
The scattering operator is defined by
\begin{eqnarray}
t(E) = H' + H' \frac{1}{E-H_0-H' } H'\,,
\end{eqnarray}
which leads to the following Lippmann-Schwinger
equation for the scattering amplitudes in each partial-wave
\begin{equation}
  t(p',p;E) = v(p',p;E)
  +\int_{C_0}dq\,q^2 
\frac{v(p',q;E)t(q,p;E)}{E-E_{1}(q)-E_{2}(q) }\,,
\label{eq:isobar-t}
\end{equation}
where the integration path $C_0$ will be specified later.
The interaction in Eq. (\ref{eq:isobar-t}) is
\begin{eqnarray}
  v(p',p;E)=\frac{g(p')g(p)}{E-M_0}\,.
\label{eq:isobar-v}
\end{eqnarray}
Eqs. (\ref{eq:isobar-t})-(\ref{eq:isobar-v}) leads to the following
well known solution
\begin{eqnarray}
  t(p',p;E) &=& \frac{g(p')g(p)}
                     {E-M_0-\Sigma(E)},
\label{eq:isobar-tij}
\end{eqnarray}
with
\begin{eqnarray}
 \Sigma(E)&=& \int_{C_0}dp\,p^2\frac{ g^2(p) }
{E-E_{1}(p)-E_{2}(p)}\,.
\label{eq:isobar-sigma}
\end{eqnarray}

From the analysis in the previous section, the
resonance poles can be found from $ t(E)= t(p_0,p_0;E)$ on the
unphysical Riemann sheet defined by $\mathrm{Im}\, p_{0} <0$ with $p_{0}$
denoting the on-shell momentum 
\begin{eqnarray}
E=\sqrt{m_1^2+p^2_{0}} + \sqrt{m_2^2+p^2_{0}}.
\label{eq:isobar-pole}
\end{eqnarray}
Obviously $p_0$ is also the pole position of the propagator in 
Eq. (\ref{eq:isobar-t}) or Eq. (\ref{eq:isobar-sigma}).

\begin{figure}
\begin{center}
\includegraphics[width=12cm]{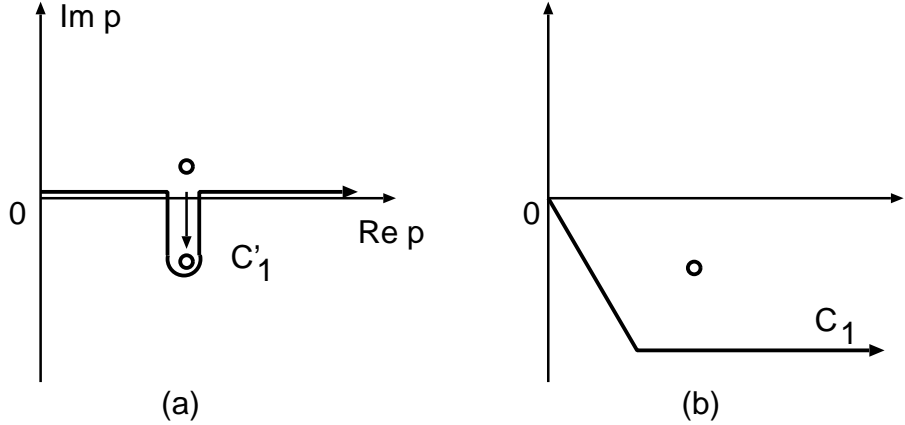}
\caption{The shift of the singularity (open circle) of the propagator
of the two-particle scattering equation Eq. (\ref{eq:isobar-t}) as energy E
moves from physical sheet to unphysical sheet. $C^\prime_1$ in (a) 
or $C_1$ in (b) is  the integration path for calculating the scattering
amplitude with E on the unphysical plane.}
\label{fig:path}
\end{center}
\end{figure}

The physical scattering amplitude at a positive energy $E$ can be obtained from
 Eq. (\ref{eq:isobar-t}) or Eq. (\ref{eq:isobar-tij}) by setting
$E \rightarrow E + i \epsilon$ with a positive $\epsilon \rightarrow 0$
and choosing the integration
contour $C_0$ to be along the real-axis of $p$ with $ 0\leq p \leq \infty$.
From Eq. (\ref{eq:isobar-t}) it is clear that
$t(E)$ has a discontinuity on the positive real $E$
\begin{eqnarray}
Dis(t(E)) & = & t(E + i \epsilon)  - t(E - i \epsilon) \nonumber \\
&=& 2\pi i \rho(p_0) v(p_0,p_0)t(E)\,,
\label{eq:t-dis}
\end{eqnarray}
where $\rho(p_0)=p_0E_1(p_0)E_2(p_0)/E$.
Thus the t-matrix has a
cut  running along the real positive $E$.
To find resonance poles, 
we need to find the solution
of Eq. (\ref{eq:isobar-t}) on the un-physical sheet with $\mathrm{Im}\, p \le 0$
on which the pole $p_0$ of the propagator 
moves into the lower p-plane, as shown in (a) of Fig.\ref{fig:path}.
From Eq. (\ref{eq:t-dis}), it is clear that the solution of
Eq. (\ref{eq:isobar-t}), with the contour $C_0$ chosen to be on the real-axis
$0\le p \le \infty$,  will
encounter the discontinuity and is not the solution
on the unphysical sheet where we want to search for the resonance poles.
It is well-known\cite{badalyan82,orlov84,pearce84,pearce89} 
that this difficulty can be overcame by
deforming the integration path to the
contour $C^\prime_{1}$  shown in (a) of Fig.\ref{fig:path}.
By this the pole will not cross the cut and the integral is
analytically continued from real positive $E$ to the  lower half of the
unphysical E-sheet with $\mathrm{Im}\, p_0 \le 0$.
Obviously, the same solution can be obtained by choosing any contour which is 
below the pole position $p_0$, such as the contour $C_1$ of
(b) of Fig.\ref{fig:path}. 

With the solution of the form of Eq. (\ref{eq:isobar-tij}),
the numerical procedure of finding resonance poles is to solve
\begin{eqnarray}
   E-M_0-\Sigma(E) = 0.
\label{eq:find-pole}
\end{eqnarray}
with 
\begin{eqnarray}
 \Sigma(E)&=& \int_{C_{1'}}dp\,p^2\frac{ g^2(p) }
{E-E_{1}(p)-E_{2}(p)} \\
&=& \int_{C_1}dp\,p^2\frac{ g^2(p) }
{E-E_{1}(p)-E_{2}(p)},
\label{eq:isobar-sigma-a}
\end{eqnarray}
for $E$ on
the unphysical Riemann sheet defined by  $\mathrm{Im}\, p_0\le 0$.
To test this numerical procedure, 
let us consider the case that $\Sigma(E)$ defined
by Eq. (\ref{eq:isobar-sigma}) can be calculated analytically.
Such an analytic form can be obtained by taking the
 non-relativistic kinematics $E_1(p)+E_2(p)=m_1+m_2+p^2/(2\mu)$
with $\mu=m_1m_2/(m_1+m_2)$ and a monopole form
for the vertex function
\begin{equation}
   g(p)=\frac{\lambda}{1+p^2/\beta^2},
\label{eq:isobar:ff1}
\end{equation}
where $\beta$ is a cut-off parameter.
The integration in $\Sigma(E)$ of Eq. (\ref{eq:isobar-sigma})
can then be done exactly to give the following simple form
\begin{eqnarray}
  \Sigma(E)&=&
              \frac{\pi\mu\beta^3\lambda^2}{2(p_0+i\beta)^2}\,,
\label{eq:sigma-ana}
\end{eqnarray}
where $p_0$ is defined by $E=m_1+m_2+p^2_0/(2\mu)$.
If the imaginary part of $p_0$ is positive (negative),
it means that we choose the poles on physical (unphysical) sheet.
Only the pole with $\mathrm{Im}\, p_0 \le0$ on the unphysical sheet is called resonance,
as discussed in the previous section.

\begin{figure}
\begin{center}
\includegraphics{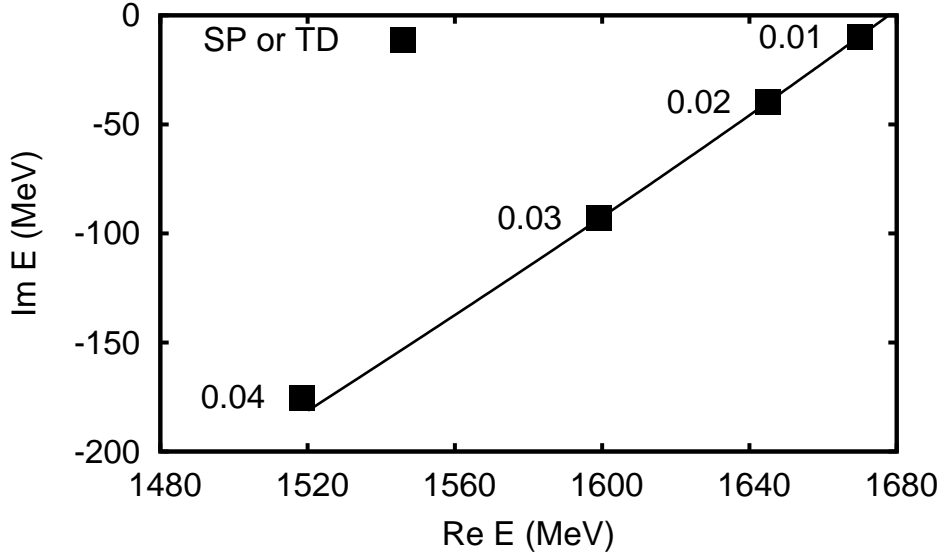}
\caption{$\lambda$ dependence of the pole position of the one-channel,
one-resonance model. 
The line  represents the pole position of the t-matrix on the 
unphysical sheet.
Solid squares  are the pole positions extracted by using the SP and TD methods.}
\label{fig:11-pole}
\end{center}
\end{figure}

The resonance poles on the unphysical
sheet ($\mathrm{Im}\, p_0 \le 0$) obtained from 
using Eq. (\ref{eq:sigma-ana}) to solve Eq. (\ref{eq:find-pole}) are 
the solid curve displayed 
in Fig.\ref{fig:11-pole} using the parameters: 
$m_{1}=m_N = 938.5 \ \mbox{MeV}$, $m_{2}=m_\pi=
139.6 \ \mbox{MeV}$, $M_0 = m_{N} + m_{\pi} + 600 \ \mbox{MeV}$ and
cutoff momentum
  $ \beta = 800 \ \mbox{MeV}$,
for a range of coupling constant
  $0 \le \lambda \le 0.04 $.
We next solve Eqs. (\ref{eq:find-pole}) and
(\ref{eq:isobar-sigma-a}) by choosing contour $C_1$ illustrated in
 Fig.\ref{fig:path}. The solutions are stable as far as the path is not too
close to the pole.  
 The solutions completely agree with the solid curve of the exact solution
and hence are not displayed  in Fig. \ref{fig:11-pole}.
Here we also note that  Eq. (\ref{eq:sigma-ana}) also allows to calculate
the poles by using the SP and TD methods.
The results for the value of $\lambda= 0.01$, $0.02$, $0.03$, $0.04$ are the 
solid squares in Fig.\ref{fig:11-pole}.
As expected we confirm the previous findings that both SP and TD 
work well for the single channel case.

\subsection{Two-channels, one-resonance}

The formula for two-channels, one-resonance case can be easily obtained by
extending the equations in the previous subsection to include channel
label $i=1,2$.  We thus have 
\begin{equation}
  t_{ij}(p',p;E) = v_{ij}(p',p;E)
  +\sum_{k}\int_{C_0}dq\,q^2
\frac{v_{ik}(p',q)t_{kj}(q,p;E)}{E-E_{k1}(q)-E_{k2}(q) }\,,
\label{eq:isobar-t2c}
\end{equation}
where $E_{kn}(p)=[m^2_{kn}+p^2]^{1/2}$ with $m_{kn}$ denoting the
mass of $n$-th particle in channel $k$, and
\begin{eqnarray}
  v_{ij}(p',p;E)=g_i(p')\frac{1}{E-M_0}g_j(p) \,.
\label{eq:isobar-v2c}
\end{eqnarray}
Eqs. (\ref{eq:isobar-t2c})-(\ref{eq:isobar-v2c}) leads to
\begin{eqnarray}
  t_{ij}(p',p;E) &=& \frac{g_i(p')g_j(p)}
                     {E-M_0-\Sigma_1(E)-\Sigma_2(E)},
\label{eq:isobar-tij2c}
\end{eqnarray}
with
\begin{eqnarray}
 \Sigma_k(E)&=& \int_{C_0}dp\,p^2\frac{ g^2_k(p) }
{E-E_{k1}(p)-E_{k2}(p)} \,.
\label{eq:isobar-sigma2c}
\end{eqnarray}
For the physical scattering amplitude at a positive energy E, 
Eq. (\ref{eq:isobar-t2c}) is solved by
setting $E \rightarrow E + i\epsilon$ with a positive $\epsilon \rightarrow 0$
and choosing $C_0$ along  the real axis $0 \le p \le \infty$.

With Eq. (\ref{eq:isobar-tij2c}), the poles 
of the scattering amplitudes are defined by
\begin{eqnarray}
E-M_0-\Sigma_1(E)-\Sigma_2(E) =0 \,.
\label{eq:find-pole2c}
\end{eqnarray}
The poles from solving the above equation can be on one of  the
four  Riemann sheets, $pp$, $up$, $uu$, and
$pu$, as explained in section III. 
The numerical procedure for finding the resonance poles on $uu$-sheet is
to solve Eq. (\ref{eq:find-pole2c}) 
for $E=E_{11}(p_{01})+E_{12}(p_{01})=E_{21}(p_{02})+E_{22}(p_{02})$
 with $\mathrm{Im}\, p_{01}\le 0$ and $\mathrm{Im}\, p_{02}\le 0$.
The integration path $C_0$ is changed  to  $C_1$
 shown in (b) of Fig.\ref{fig:path} to calculate both self-energies
$\Sigma_1(E)$ and $\Sigma_2(E)$ of Eq. (\ref{eq:isobar-sigma2c}). 
Of course the contour $C_1$ for the integration over the momentum for $i$-th 
channel must be below the pole $p_{0i}$ defined by 
$E-E_{i1}(p_{0i})- E_{i2}(p_{0i}) =0$.
Here we note that for
finding the poles on $pu$-sheet ($up$-sheet),
the contour $C_0$ is replaced by $C_1$ only for $\Sigma_2(E)$ ($\Sigma_1(E)$).

\begin{figure}
\begin{center}
\includegraphics{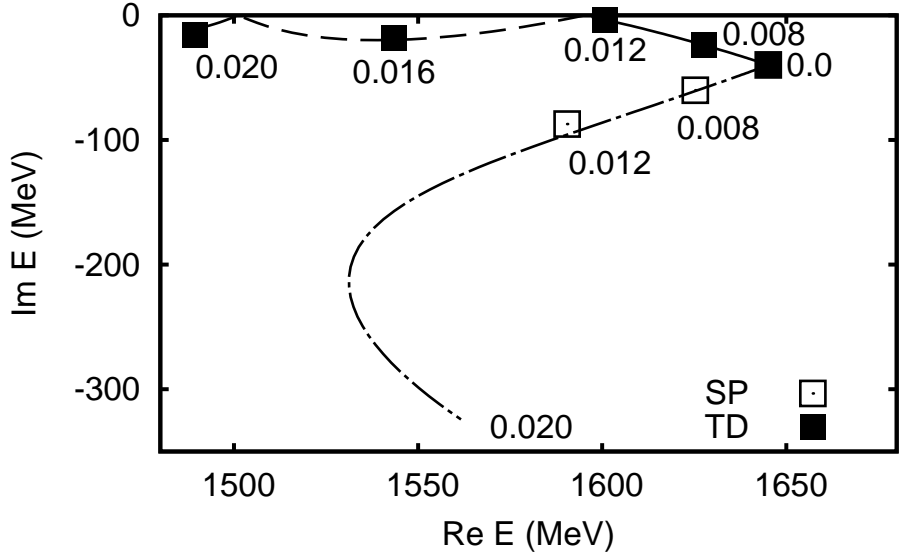}
\caption{$\lambda_2$ dependence of the pole positions on the
$uu$-sheet (dash-dotted curve),  $up$-sheet (solid curve) 
and $pu$-sheet (dashed curve) of
the two-channels, one-resonance model. 
The open squares (solid squares) are obtained from using the SP (TD) method.}
\label{21-pole}
\end{center}
\end{figure}

To test numerical procedures and to further test the SP and TD methods, let us 
consider again the
non-relativistic kinematics $E_{i1}(p)+E_{i2}(p)= \Theta_i +p^2/(2\mu_i)$ 
with $\Theta_i=m_{i1}+m_{i2}$ and
$\mu_i=m_{i1}m_{i2}/(m_{i1}+m_{i2})$. 
This will allow us to find the exact solutions by choosing the monopole 
form factor
\begin{eqnarray}
   g_i(p)=\frac{\lambda_i}{1+p^2/\beta_i^2}\,.
\label{eq:isobar:ff}
\end{eqnarray}
We then have
\begin{eqnarray}
  \Sigma_i(E)&=&
              \frac{\pi\mu_i\beta_i^3\lambda_i^2}{2(p_i+i\beta_i)^2}.
\label{eq:sigma-ana2c-1}
\end{eqnarray}
With Eq. (\ref{eq:sigma-ana2c-1}), the poles defined by 
Eq. (\ref{eq:find-pole2c}) can be found by solving algebraic equations.
For numerical calculations, we consider a case similar to $\pi N$ scattering in $S_{11}$
partial wave:
(1)channel-1 is $\pi N$ with $m_{11}=m_\pi=139.6$ MeV ,
$m_{12}=m_N=938.5$ MeV, and
$\beta_1 = 800 \ \mbox{MeV} $, (2)channel-2 is $\eta N$ with 
$m_{21}=m_\eta=547.45$ MeV ,
$m_{22}=m_N=938.5$ MeV, and $\beta_2 = 800 \ \mbox{MeV}$, (3)
bare mass $M_0 = m_{\pi} + m_{N} + 600 \ \mbox{MeV}$.
The results are shown 
in Fig.\ref{21-pole}. The dash-dotted (solid) curves are the calculated
poles on $uu$-sheet ($up$-sheet) with the coupling
constants $  \lambda_1 = 0.02$ for $N_0 \rightarrow \pi N$ 
 and a range of
$  \lambda_2 = 0 - 0.02$ for $N_0 \rightarrow \eta N$.
We see that when $\lambda_2$ is 0, which is the single-channel case, 
the $uu$-pole
and the $up$-pole are on the same position. They then split as $\lambda_2$
increases. 

 We next evaluate Eq. (\ref{eq:isobar-sigma2c}) for $E$ on the
 $uu$-sheet, $up$-sheet
or $pu$-sheet by appropriately
choosing the path $C_0$, as described above.
The  poles are found when the
calculated $\Sigma_1(E)$ and $\Sigma_2(E)$  satisfy
Eq. (\ref{eq:find-pole2c}). We find that  the poles obtained by this
numerical procedure reproduce accurately the dash-dotted($uu$-sheet),
solid($up$-sheet) and dashed($pu$-sheet)
curve in Fig.\ref{21-pole} and hence are omitted there.
 Thus this analytic continuation method can be used in practice
to find the resonance
poles (i.e. poles on $uu$-sheet as defined in section III) for general
case that $\Sigma_i(E)$ can not be integrated analytically.

We now turn to examining the SP and TD methods.
In Fig.\ref{21-pole}, we show that  the SP method (open squares) reproduces 
the poles (dash-dotted curve) on the $uu$-sheet
only at $\lambda_2 \le 0.015$. 
When $\lambda_2$ is larger than 0.015 where the magnitude of $\mathrm{Im}\, E_R$
continues to increase, the speed has no maximum and the SP method
fails to find the pole. 
This is another example showing that SP method has its limitation.
On the other hand, the TD method
(solid squares) can reproduce  the poles on the $up$-sheet (solid curve) 
and $pu$-sheet (dashed curve)
in the considered range of parameters. Here we see that the SP and TD find 
different poles which have different physical meanings in the Hamiltonian
formulation. One can show\cite{gw}
 that the poles on $uu$-sheet are due to the process that
an unstable
system is created and then decays during the collision and are called the
resonance poles.  
The physical interpretations of the poles on $pu$ and $up$-sheets remain to be
developed.

It is interesting to point out here
 that the two-channel Briet-Wigner form analyzed
 in detail in the previous  section can be derived 
from the two-channels, one-resonance
model if the non-relativistic kinematics is used.
To see this, we first write
the non-relativistic relation between the S-matrix and the T-matrix
\begin{eqnarray}
  S_{ij}(E)&=&\delta_{ij}+2iT_{ij}(E),
\label{eq:s-matrix}\\
  T_{ij}(E)&=&-\pi\sqrt{\mu_{i}p_{i}\mu_{j}p_{j}}
  t_{ij}(p_{i},p_{j};E),
\label{eq:t-matrix}
\end{eqnarray}
where $p_i$ is the on-shell momentum in channel $i$
\begin{equation}
  p_{i}=\sqrt{2\mu_{i}(E-\Theta_{i})}.
\end{equation}
With the above and the analytic form Eq. (\ref{eq:sigma-ana2c-1}) for
$\Sigma_{i}(E)$ , we can write the $1 \rightarrow 1$ elastic
scattering amplitude of
Eq. (\ref{eq:t-matrix}) as
\begin{eqnarray}
T_{11}(p_1,p_1,E)= \frac{-p_1\gamma_1(p_1)}
{E-M(E)+ip_1\gamma_1(p_1)+ip_2\gamma_2(p_2)},
\label{eq:T11}
\end{eqnarray}
where $\gamma_i(p_i)=\pi\mu_i g_i^2(p_i) > 0$ and
\begin{eqnarray}
M(E) &=& M^0 +\sum_{k}P\int p^2 dp
\frac{ g^2_k(p) }{E-m_{k1}-m_{k2}-\frac{p^2}{2\mu_k}}
\end{eqnarray}
where $P$ means taking the principal-value integration.
By using Eq. (\ref{eq:s-matrix}), we then have the $1\rightarrow 1$
elastic part of the S-matrix
\begin{eqnarray}
S_{11} & = & \frac{E - M(E) - i p_1 \gamma_1(p_1) + i p_2 \gamma_2(p_2)}
                  {E - M(E) + i p_1 \gamma_1(p_1) + i p_2 \gamma_2(p_2)}.
\label{eq:S11}
\end{eqnarray}
If the E-dependence of $M(E)$ and $\gamma_i(p_i)$
are further neglected, 
Eqs. (\ref{eq:T11}) and
(\ref{eq:S11}) are identical to what are usually called the two-channel
Breit-Wigner resonant amplitude discussed in the previous section.
Thus the conditions under which SP and TD are valid can be related now
to the parameters of the vertex function 
$g_i(p)$ within this two-channels, one-resonance model.

\subsection{Two-channels, two-resonances}

For the two-channels, two-resonances case, 
the scattering amplitude is defined by
the same Eq. (\ref{eq:isobar-t2c}), but with the following driving term
\begin{equation}
  v_{ij}(p',p;E) = g_{i1}(p')\frac{1}{E-M_1}g_{j1}(p)
+g_{i2}(p')\frac{1}{E-M_2}g_{j2}(p).
\end{equation}
The scattering amplitude is then of the following form
\begin{eqnarray}
  t_{ij}(p',p;E) =
 \sum_{\alpha,\beta}
g_{i,\alpha}(p')[D^{-1}(E)]_{\alpha,\beta} g_{j,\beta}(p).
\label{eq:2c2r-t}
\end{eqnarray}
The propagator  $D(E)$ in Eq. (\ref{eq:2c2r-t}) is
\begin{eqnarray}
 [ D(E)]_{\alpha,\beta} = [E-M_\alpha]\delta_{\alpha,\beta}
-\Sigma_{\alpha,\beta}(E),
\label{eq:2c2r-resp}
\end{eqnarray}
with
\begin{eqnarray}
  \Sigma_{\alpha,\beta}(E) = \sum_{i}\int_{C_0}dq\,q^2
 \frac{g_{i,\alpha}(q)g_{i,\beta}(q)}{E-E_{i1}(q)-E_{i2}(q)
+i\varepsilon}.
\label{eq:2c2r-sigma}
\end{eqnarray}
The poles are defined by 
\begin{eqnarray}
\mathrm{Det}\, D(E) &=& [E-M_1-\Sigma_{11}(E)][E-M_2-\Sigma_{22}(E)]
-\Sigma_{12}(E)\Sigma_{21}(E)\nonumber \\
& =&0\,.
\label{eq:2c2r-pole}
\end{eqnarray}

The numerical procedures of finding the resonance poles on the $uu$-sheet
from Eqs. (\ref{eq:2c2r-sigma}) and (\ref{eq:2c2r-pole}) are 
the same as that in the previous two subsections. Namely the
path $C_0$ of Eq. (\ref{eq:2c2r-sigma}) is set to be the path
$C_1$ shown in Fig.\ref{fig:path}
in evaluating the integrals for $E$ on the $uu$-sheet where the on-shell
momenta are $\mathrm{Im}\, p_i < 0$ for $i=1,2$ channels. To test this, we again
choose the non-relativistic kinematics and the  monopole form
factor like Eq. (\ref{eq:isobar:ff}). The self energy $\Sigma_{\alpha,\beta}(E)$
then takes the analytic form similar to Eq. (\ref{eq:sigma-ana}).
 The condition Eq. (\ref{eq:2c2r-pole}) 
can then be expressed in a analytic form from which
the pole positions on the unphysical sheet can be easily obtained. 

We only state that the resulting poles on the unphysical sheets are 
reproduced by the numerical analytic continuation method described above.
Instead our focus here is to further test the SP and TD  methods
for the situation that two resonances are  close and could overlap.
We again consider $\pi N$ and $\eta N$ channels and use 
the following form factor
\begin{equation}
   g_{i\alpha}(p)=\frac{\lambda_{i\alpha}}{1+p^2/\beta_{i\alpha}^2},
\end{equation}
where $\alpha = 1,2$ denote the $\alpha$-th bare state with mass $M_\alpha$.
The four cut-off parameter $\beta_{i\alpha}$ and four coupling
constants $\lambda_{i\alpha}$ are taken to be:
$\beta_{11}=\beta_{12}=\beta_{21}=\beta_{22}=800$ MeV,
 $\lambda_{11}=0.005$, $\lambda_{12}=0.01$, $\lambda_{21}=0.003$ and 
$\lambda_{22}=0.008$.
One of the bare masses $M_1$ is fixed in the calculations.
$M_2$ is defined by
\begin{eqnarray}
M_2 = M_{\pi}+M_N+\tilde{M}_2,
\end{eqnarray}
where $\tilde{M}_2$ is varied for examining how the poles move as $M_2$
moves away from the $M_1=M_{\pi}+M_{N}+550 = 1628$ MeV.

\begin{figure}
\begin{center}
\includegraphics{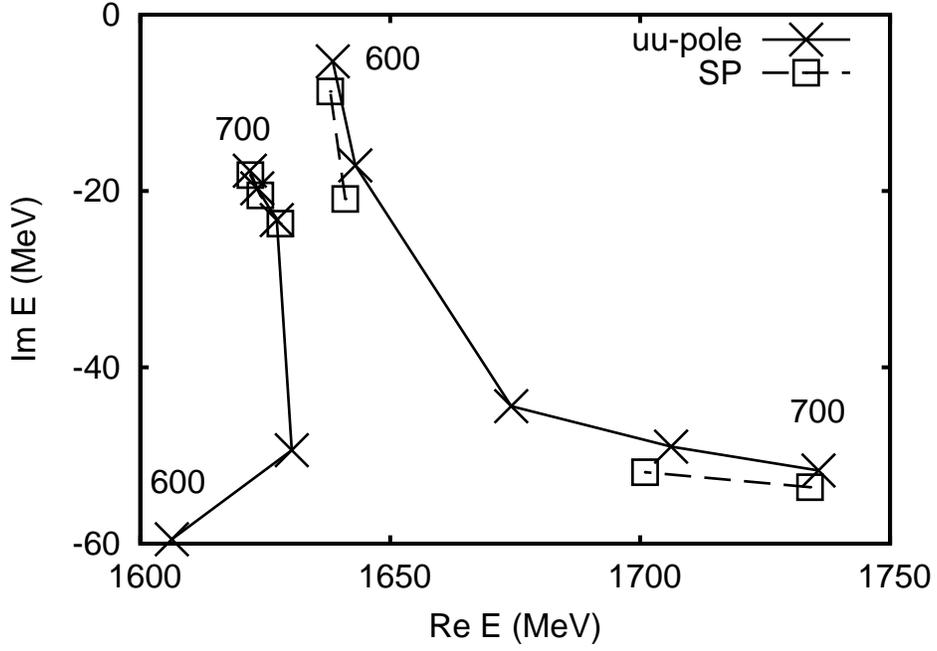}
\caption{
Pole positions (crosses connected by solid lines)
on the $uu$-sheet of the two-channels, two-resonances model.
The results from using the SP method are the open squares connected by
the dashed lines. The numbers on the figure are the value of $\tilde{M}_2$.
}
\label{pole3rd-sp}
\end{center}
\end{figure}

\begin{figure}
\begin{center}
\includegraphics{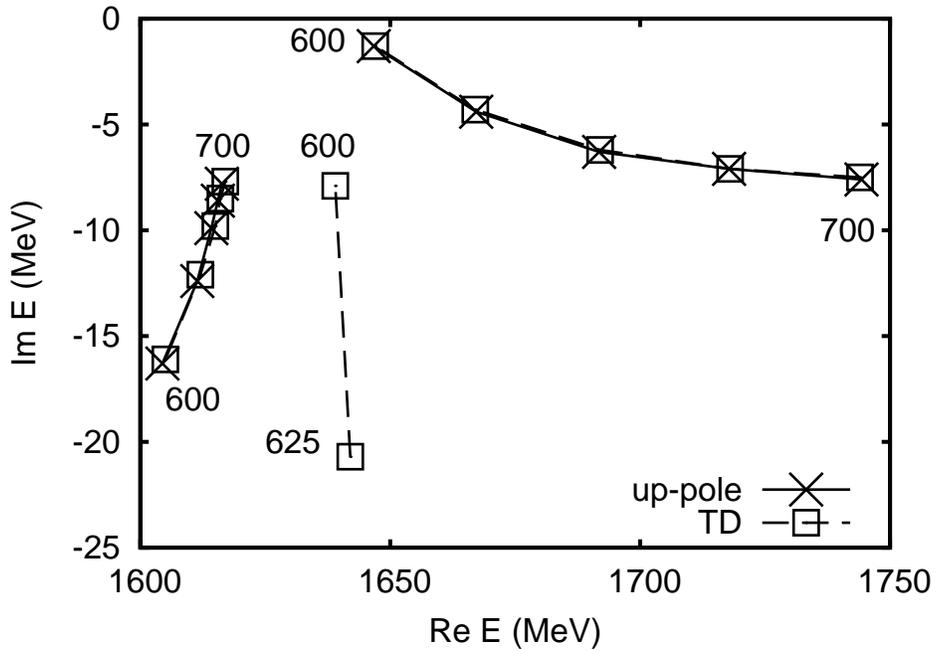}
\caption{
Pole positions (crosses connected by solid lines)
on the $up$-sheet of the two-channels, two-resonances model. 
The results from using the TD method are the open squares connected by 
the dashed lines. The numbers on the figure are the value of $\tilde{M}_2$.
}

\label{pole2nd-td}
\end{center}
\end{figure}

The pole positions are searched numerically by
using the analytic continuation method described above.
There are two poles on the $uu$-sheet and the other two  on the $up$-sheet.
As $\tilde{M}_2$ varies, these two poles will develop
two trajectories. They are 
the crosses connected by the solid curves shown in
Fig.\ref{pole3rd-sp} for $uu$-sheet and Fig.\ref{pole2nd-td} for the 
$up$-sheet. According to the findings we made in section III, 
the poles, $(M_R,-i \Gamma_R/2)$, found by the SP (TD)  method
 should be compared with 
the poles on $uu$ ($up$) sheet of
Fig. \ref{pole3rd-sp} (\ref{pole2nd-td}). We now discuss these two comparisons.

We see from Fig. \ref{pole3rd-sp} that 
in the regions near $\tilde{M}_2 = 700$ MeV, the positions 
($\mathrm{Re}\, E$) of
these two poles are far from each other
and we find that SP (open squares connected by dashed lines near
the point marked 700) works well.
When $\tilde{M}_2$ is reduced to $600$ MeV where the positions
($\mathrm{Re}\, E$)
of two  poles
move closer, the  SP  method can find only one pole
( open squares connected by dashed line) near the
top end of the trajectory on the right hand side.
Apart from the points on the dashed lines, SP method fails to find  poles
close to the poles 
on the solid curves which are obtained numerically
by the analytic continuation method.

The results for examining TD is shown in Fig.\ref{pole2nd-td}.
We see that  TD can find two poles
 on the $up$-sheet in the
considered range of $\tilde{M}_2 = 600 - 700$. The results
are the open squares connected by dashed lines which are indistinguishable
from the crosses connected by  solid curves which were obtained by analytic continuation method.
But TD  obtains another two poles at $\tilde{M}_2 = 600$ and 
$\tilde{M}_2 = 625$, as indicated by the dashed line in the middle of
Fig.\ref{pole2nd-td}. 
 The positions of these two poles are very close to those
obtained from SP and they are interpreted as poles on $uu$-sheet.
As we have discussed in section II, TD is sensitive to both zero and
pole of the S-matrix on $pp$ and $uu$-sheets. 
In this example, width of the poles on
$uu$-sheet become comparable to the poles on $up$-sheet(zero on physical
sheet) and hence TD could find pole on $uu$-sheet.
The results shown in Fig. \ref{pole3rd-sp} and Fig. \ref{pole2nd-td}
further indicate the limitation of SP and TD methods.

The results from the above several models have shown 
that the TD method based on Eq. (\ref{eq:td-kel-1}),
where the phase of the elastic channel is used instead of  the
eigen phases discussed in
 Refs. \cite{hazi,iga04,haber},
gives both the resonance poles on the $uu$-sheet and the zeros(shadow poles).
Our findings could provide some information for investigating the
differences between Refs.\cite{haber} and\cite{kel2008}.
We also want to mention here that an improved SP method
using higher order derivatives of the amplitudes was proposed in
 Ref. \cite{ceci2008}.  
It may be interesting to compare this method with the TD method.
While it could be interesting to address the questions concerning 
these recent developments,  they are far from the main focus of this
paper and will not be discussed further.  

\section{Analytic continuation of resonance model with unstable
particle channels}

For meson-baryon reactions, the nucleon resonances can decay into
some unstable particle channels such as the $\pi \Delta$, $\rho N$, $\sigma N$
considered in the model of Ref.\cite{msl}. Here we discuss 
the analytic continuation
method to find resonance poles within such a reaction model. 

It is sufficient to consider the one-channel and one resonance case. 
The scattering formula is then identical to that presented in subsection IV.A.
The only difference is that  one of the particles in the open
channel can further
decay into a two particle state. To be specific, let us consider
the $\pi \Delta$ channel. Within the same Hamiltonian formulation\cite{msl} 
used in the previous section,  the scattering amplitude can then be written as
\begin{eqnarray}
t(p',p,E)=
\frac{g_{N^*,\pi\Delta}(p')g_{N^*,\pi\Delta}(p)}{E-M_0 -\Sigma_{\pi \Delta}(E)}
\label{eq:unst-t}
\end{eqnarray}
with 
\begin{eqnarray}
\Sigma_{\pi \Delta}(E) =\int_{C_2}p^2\, dp\frac{g^2_{N^*,\pi\Delta}(p)}
{E-E_\pi(p)-E_\Delta(p)-\Sigma_\Delta(p,E)}
\label{eq:unst-sigma-pid}
\end{eqnarray}
where
\begin{eqnarray}
\Sigma_\Delta(p,E) = \int_{C_3} q^2 dq \frac{g^2_{\Delta,\pi N}(q)}
{E-E_\pi(p)- [(E_\pi(q)+E_N(q))^2+p^2]^{1/2}}\, .
\label{eq:unst-sigma-d}
\end{eqnarray}

\begin{figure}
\begin{center}
\includegraphics{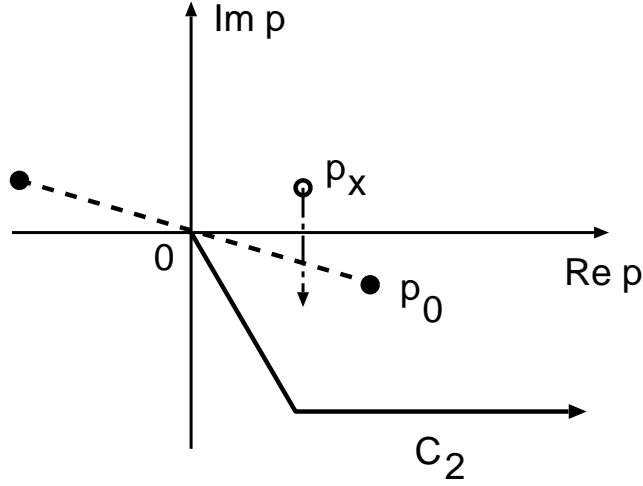}
\caption{ Contour $C_2$ for calculating the $\pi\Delta$ self energy
on unphysical sheet. See the text for the explanations of the dashed line and
the singularity $p_x$.}
\label{fig:c2}
\end{center}
\end{figure}

To obtain the $\pi\Delta$ self energy for complex $E$,
the analytic structure of the integrand of Eq. (\ref{eq:unst-sigma-pid})
should be examined first.
The discontinuity of the $\pi\Delta$ propagator in the integrand of 
Eq. (\ref{eq:unst-sigma-pid}) is the $\pi\pi N$ cut 
along the real axis between $\pm p_0$ ($-p_0 \le p \le p_0$)
which is obtained by solving
\begin{eqnarray}
E=E_\pi(p_0)+[(m_\pi+m_N)^2+p^2_0]^{1/2}.
\end{eqnarray}
For finding the resonance poles on $uu$-sheet  with $\mathrm{Im}\, p_0 \le 0$, 
the integration contour $C_2$ of Eq. (\ref{eq:unst-sigma-pid})
 must be chosen below this cut which is the dashed line in 
Fig.\ref{fig:c2}.
There is also a singularity in the integrand of Eq. (\ref{eq:unst-sigma-pid}) 
at momentum $p=p_x$, which satisfies
\begin{eqnarray}
E-E_\pi(p_x)-E_\Delta(p_x)-\Sigma_\Delta(p_x,E)=0.
\label{eq:px}
\end{eqnarray}
Physically, this singularity corresponds to the $\pi\Delta$ two-body 
scattering state.
For $E$ with large imaginary part,
$p_x$ can be below the $\pi\pi N$ cut as also indicated in
Fig.\ref{fig:c2}. 
Therefore the integration contour of momentum $p$
must be chosen to be below the $\pi\pi N$ cut (dashed line)
 and the singularity $p_x$, such
as the contour $C_2$ shown in  Fig.\ref{fig:c2}.

\begin{figure}
\begin{center}
\includegraphics{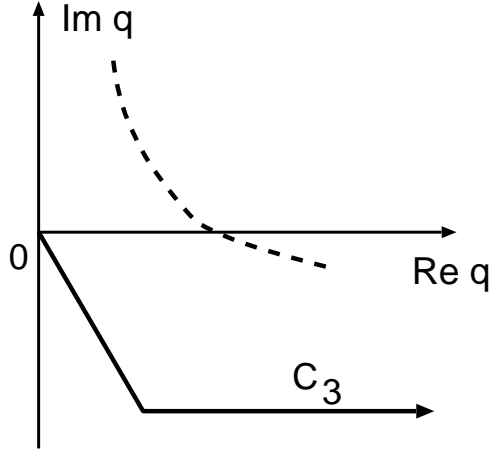}
\caption{ Contour $C_3$ for calculating the $\pi N$ self energy
on the unphysical sheet. Dashed curve is the singularity $q_0$ of
the propagator in Eq. (\ref{eq:unst-sigma-d}), which depends on
the spectator momentum $p$ on the contour $C_2$ of Fig.\ref{fig:c2}. }
\label{fig:c3}
\end{center}
\end{figure}

The singularity position $q_0$ 
of the propagator in Eq. (\ref{eq:unst-sigma-d}) depends on 
spectator momentum $p$
\begin{eqnarray}
E-E_\pi(p) = [(E_\pi(q_0)+E_N(q_0))^2+p^2]^{1/2} \,.
\end{eqnarray}
Therefore the singularity $q_0$
 moves along the dashed curve  in Fig.\ref{fig:c3}
when the momentum $p$ varies along the path $C_2$ of Fig.\ref{fig:c2}.
To analytically continue $\Sigma_\Delta(p,E)$ from positive energy $E$ to
the un-physical plane with $\mathrm{Im}\, p\le 0$, we need to choose the contour $C_3$
of Eq. (\ref{eq:unst-sigma-d}) which must be
 below $q_0$. A possible contour $C_3$ is the solid curve in Fig.\ref{fig:c3}.

To verify the numerical procedures described above, we again consider
 non-relativistic  kinematics and monopole form factor.
With the similar analytic form Eq. (\ref{eq:sigma-ana}), we have
\begin{eqnarray}
\Sigma_\Delta(p,E)=
\frac{\pi \mu_{\pi N}g^2_{\Delta,\pi N}\beta^3_{\Delta,\pi N}}
{2(\bar{k}+i\beta_{\Delta,\pi N})^2}
\label{eq:unst-sigma-da}
\end{eqnarray}
where
\begin{eqnarray}
\bar{k}=\left[2\mu_{\pi N} \left(E-2m_\pi - m_N - \frac{p^2}{2\mu_{\pi\pi N}}\right)\right]^{1/2},
\end{eqnarray}
with $\mu_{\pi\pi N}=m_\pi(m_\pi+m_N)/(2 m_\pi + m_N)$.
With Eq. (\ref{eq:unst-sigma-da}), we can solve Eq. (\ref{eq:px}) and verify
its relation with $\pi\pi N$ cut as discussed above and illustrated in
Fig.\ref{fig:c2}. Eq. (\ref{eq:unst-sigma-da}) and the chosen monopole 
form factor also allow us to get
\begin{eqnarray}
\Sigma_{\pi\Delta} =
\int_{c_2} p^2 dp\frac{g^2_{N^*,\pi \Delta}}{(1+p^2/\beta^2_{N^*,\pi \Delta})^2}
\frac{1}{D_{\pi \Delta}(p,E)}
\label{eq:unst-sigma-pida}
\end{eqnarray}
with
\begin{eqnarray}
D_{\pi \Delta}(p,E) = E-m_\pi - m_\Delta - \frac{p^2}{2\mu_{\pi \Delta}}
-\frac{\pi \mu_{\pi N}g^2_{\Delta,\pi N}\beta^3_{\Delta,\pi N}}
{2(\bar{k}+i\beta_{\Delta,\pi N})^2}.
\label{eq:d-pid}
\end{eqnarray}
Unfortunately, Eq. (\ref{eq:unst-sigma-pida}) can not be integrated out
 analytically
for directly checking our numerical procedure for searching resonance poles.

\begin{figure}
\begin{center}
\includegraphics[width=8cm]{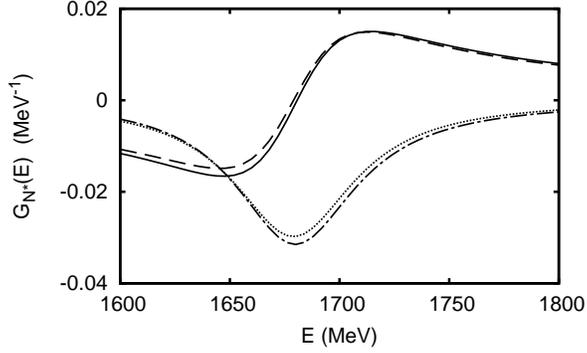}
\caption{$N^*$ Green function.
 The solid (dash-dotted) curve is the real (imaginary) part 
of exact Green function $G_{N^*}(E)=1/(E-M_0-\Sigma_{\pi\Delta}(E))$.
They are compared with the dashed (dotted) curve
of the real (imaginary) part of the approximate Green function
$G_{N^*}^{approx}$. }
\label{fig:g-pid}
\end{center}
\end{figure}

We test our analytic continuation method  by the following procedure.
We calculate Eq. (\ref{eq:unst-sigma-pida}) numerically to find the
pole position $E_{R}$ by solving
$E_R-M_0-\Sigma_{\pi \Delta}(E_R)=0$ of the denominator of
Eq. (\ref{eq:unst-t}).
With the parameters:
\begin{eqnarray}
\beta_{N^*,\pi \Delta}  = & 800MeV\,\,,  \ g_{N^*,\pi\Delta}=& 0.02MeV^{-1/2}, 
\nonumber \\
\beta_{\Delta,\pi N}    = & 200MeV\,\,,  \ g_{\Delta,\pi N} =& 0.05 MeV^{-1/2},
\nonumber \\
  M_0= & 650MeV+m_\pi + m_N & \nonumber,
\end{eqnarray}
we find $E_{R}=( 1679.1, - 33.6i) $ MeV. 
We then construct an approximate propagator
\begin{eqnarray}
G^{approx}_{N^*}(E)=\frac{1}{E-E_{R}}\,.
\label{eq:g-appr}
\end{eqnarray}
For the positive E, we find that
$G^{approx}_{N^*}(E)$ is in good agreement with the direct calculation of
$G_{N^*}(E)=1/(E-M_0-\Sigma_{\pi \Delta}(E))$ by 
using Eq. (\ref{eq:unst-sigma-pida}).
The results are shown in Fig.\ref{fig:g-pid}.
It is clear that the resonance pole found by
our  analytic continuation method
can reproduce what is expected for a resonance propagator for real positive E.
In this way our numerical procedure is justified and can be applied
to solve Eqs. (\ref{eq:unst-t}) - (\ref{eq:unst-sigma-d}).

\section{Resonance model with non-resonant interactions}

With the numerical methods described above, we can proceed to extract
the resonance poles within a coupled-channels 
model which also include non-resonant interactions.
In this section, we explain 
how this can be done for 
the  $\pi N$ model developed in Refs.\cite{msl,jlms}.

Recalling the formulations presented in Refs.\cite{msl,jlms}, the
t-matrix considered is of the following form
\begin{equation}
  t_{ij}(p',p;W) = 
\bar{t}_{ij}(p',p;W) + \sum_{\alpha,\beta}
\bar{\Gamma}_{i,\alpha}(p';W)[D^{-1}(W)]_{\alpha,\beta}
\bar{\Gamma}_{j,\beta}(p;W).
\label{eq:msl-t}
\end{equation}
where $i,j$ can be stable channels $ \pi N$ and $ \eta N$, or 
unstable channels $ \pi \Delta, \rho N$, and $\sigma N$, and $\alpha$
denotes a bare resonant state with a mass $M_\alpha$.
 
For extracting resonance poles, we now apply the methods presented 
in previous sections to choose appropriate contours for calculating
various integrations on unphysical E-plane with $\mathrm{Im}\, p\le 0$.
The non-resonant t-matrix $\bar{t}_{ij}(p',p;E)$ is defined by the 
following couple-channel equations with the non-resonant
potential $v_{ij}(p',p)$,
\begin{equation}
  \bar{t}_{ij}(p',p;E) = v_{ij}(p',p)
  +\sum_{k}\int_{C_4}dq\,q^2 
  \frac{v_{ik}(p',q)\bar{t}_{kj}(q,p;E)}{E-E_k(q)-\Sigma_k(q,E)+i\varepsilon},
\label{eq:msl-nont}
\end{equation}
where the contour is $C_4=C_1$ ((b) of Fig.\ref{fig:path})
for $k = \pi N, \eta N$, and $C_4=C_2$ (Fig.\ref{fig:c2}) for
$k = \pi \Delta, \rho N, \sigma N$, and
\begin{equation}
  E_k(p) = \sqrt{m_{k1}^2+p^2}+\sqrt{m_{k2}^2+p^2}.
\end{equation}
The self-energies in Eq. (\ref{eq:msl-nont})  are $
\Sigma_{\pi N} (q,E)= \Sigma_{\eta N}(q,E)=0$ and $\Sigma_{\pi \Delta} (q,E)$,
$ \Sigma_{\rho N}(p,E)$, and $\Sigma_{\sigma N}(q,E)$ are defined by the
same Eq. (\ref{eq:unst-sigma-pid}) with appropriate changes of mass parameters
and the choice of contour $C_2$ and $C_3$  shown in Figs. \ref{fig:c2}
and \ref{fig:c3}.

The dressed vertex  $\bar{\Gamma}_{i}(p;W)$ is 
determined by the bare vertex $\Gamma_{i}(p)$ and the meson-baryon loop,
\begin{equation}
  \bar{\Gamma}_{i,\alpha}(p;E) = \Gamma_{i,\alpha}(p)
  +\sum_{k}\int_{C4} dq q^2 
  \frac{\bar{t}_{ik}(p,q;E)\Gamma_{k}(q;E)}
{E-E_k(q)-\Sigma_{k}(p,E)+i\varepsilon}.
\label{eq:dress-v}
\end{equation}
The resonance propagator  $D(E)$ in Eq. (\ref{eq:msl-t}) is
\begin{equation}
 [ D(E)]_{\alpha,\beta} = [E-M_\alpha]\delta_{\alpha,\beta}
-\Sigma_{\alpha,\beta}(E),
\label{eq:msl-resp}
\end{equation}
with
\begin{equation}
  \Sigma_{\alpha,\beta}(E) = \sum_{k}\int_{C_4}dq\,q^2 
  \frac{\Gamma_{k,\alpha}(q)\bar{\Gamma}_{k,\beta}(q;E)}{E-E_k(q)-\Sigma_k(q,E)
+i\varepsilon}.
\end{equation}
In Ref.\cite{jlms}, the above equations are solved on the real-axis by
using the standard method of subtraction. Here we solve the equations by
choosing contours indicated above. We first verify that our numerical results
obtained here for positive real E agree with that of Ref.\cite{jlms}.
This establish our numerical procedure in this complex
five-channel model.

Here we show the results for some of the 
$S$, $P$, $D$, and $F$ partial waves of 
$\pi N$ scattering within the model of Ref.\cite{jlms}.
We search the resonance poles by looking for zeros of 
the resonant propagator $D(E)$ defined by Eq. (\ref{eq:msl-resp}).
Our results from using the analytic continuation method
are shown in the second column of Table \ref{tab:ress11s31}. They are 
compared with those extracted by using SP and TD methods described in
the previous sections 
 as well as the values listed by PDG. 
The Breit-Wigner resonance parameters are also given by PDG, but
are not considered here.
We see in Table \ref{tab:ress11s31} that the SP method fails to
find the first resonance pole in the $S_{11}$ partial wave.
We also see that  while the
real parts of the resonance poles from different approaches are within the
ranges of PDG, the extracted imaginary parts can differ by as much as a 
factor of two or three. 

The model of Ref.\cite{jlms} is currently
being improved by also fitting other data of $\pi N$ and $\gamma N$ reactions.
For example, some progress has been made to also fit
the data of
$\pi N \rightarrow \pi\pi N$\cite{kjlms08}, 
$\gamma N \rightarrow \pi N$\cite{jlmss08}, and
$\pi N \rightarrow \eta N$\cite{djlss08}.
We thus do not include the results for other partial waves
in Table \ref{tab:ress11s31}. Our purpose here is to simply demonstrate
how the analytic continuation method
works for a realistic model.
The full resonance parameters, 
including the extracted residues and the
relations to the Breit-Wigner parameters listed by PDG, 
extracted from our complete analysis of all 
$\pi N \gamma, N \rightarrow \pi N, \eta N, \pi\pi N$ will be reported 
elasewhere.

\begin{table}
\caption{Resonance poles extracted from the $\pi N$
scattering amplitudes of Ref.\cite{jlms}}
\begin{tabular}{c|cc|cc|cc|cc} \hline\hline
 & Analytic Continuation & & Speed PLot & & Time Delay & & PDG & \\ \hline
 & Re & Im & Re & Im & Re & Im & Re & Im \\ \hline
S11 & 1540 & -191 & - & -  & 1543 & -52 & 1490 $\sim$ 1530
  & -45 $\sim$ -125 \\
  & 1642 & -41 & 1644 & -89 &   1645 & -61 & 1640 $\sim$ 1670
  & -75 $\sim$  -90 \\ \hline
S31 & 1563 & -95 & 1574 & -67 & 1616 & -53 & 1590 $\sim $ 1610
    & -57 $\sim $-60 \\ \hline
P33 & 1211 & -50 & 1212 & -49 & 1212 & -49 & 1209 $\sim $ 1211
    & -49 $\sim $-51 \\ \hline
D13 & 1521 & -58 & 1525 & -57 & 1522 & -11 & 1505 $\sim $ 1515
    & -52 $\sim $-60 \\ \hline
F15 & 1674 & -53 & 1671 & -59 & 1683 & -24 & 1665 $\sim $ 1680
    & -55 $\sim $-68 \\ \hline
\end{tabular}
\label{tab:ress11s31}
\end{table}

We now turn to discussing whether the extracted resonance pole $M$
can be used to evaluate the $N^*$ propagator defined by Eq. (\ref{eq:msl-resp}) for the physical positive E.  
Let us consider the $S_{31}$ case listed in Table \ref{tab:ress11s31}.
Its $N^*$ propagator $G_{N^*}(E)$ can be written as
\begin{eqnarray}
 G_{N^*}(E)&=&\frac{1}{E-M_0-\Sigma(E)}, 
\label{eq:nstar-g-exact}\\
\Sigma(E)&=&
\Sigma_{\pi N}(E)+\Sigma_{\pi\Delta}(E)+\Sigma_{\rho N}(E).
\end{eqnarray}
By using the analytic continuation methods described in the previous
sections, the  resonance energy $M = (1563 - i 95) $ MeV
is found numerically by solving
\begin{eqnarray}
 M-M_0-\Sigma(M)=0.
\end{eqnarray}

We now perform the Laurent expansion of $G_{N^*}(E)$ for real $E$
around the pole position $M$ 
\begin{eqnarray}
 G_{N^*}(E)&=&[
(1-\Sigma'(M))(E-M)-\frac{1}{2}\Sigma''(M)(E-M)^2+\cdots]^{-1} \nonumber \\
&=&\frac{1}{E-M}\left [ \frac{1}{1-\Sigma'(M)
-\frac{1}{2}\Sigma''(M)(E-M) + \cdots} \right ] \nonumber \\
&=&\frac{1}{E-M}\left[ \frac{1}{1-\Sigma'(M)}
+\frac{\Sigma''(M)(E-M)}{2(1-\Sigma'(E))^2} + \cdots \right] \nonumber \\
&=&\frac{1}{E-M}\cdot\frac{1}{1-\Sigma'(M)}
+\frac{\Sigma''(M)}{2(1-\Sigma'(M))^2} + \cdots.
\label{eq:nstar-g}
\end{eqnarray}
The naive $1/(E - M)$ works well for a model studied in the previous section.
However when the enery dependence of the self energy $\Sigma$ 
becomes large, two important modifications should be considered:
(1) at the pole, the residue is not one
but modified by the field renormalization factor $Z=1/(1-\Sigma'(M))$.
(2)The second term of the last expression gives a constant term.

\begin{figure}
\begin{center}
\includegraphics[width=8cm]{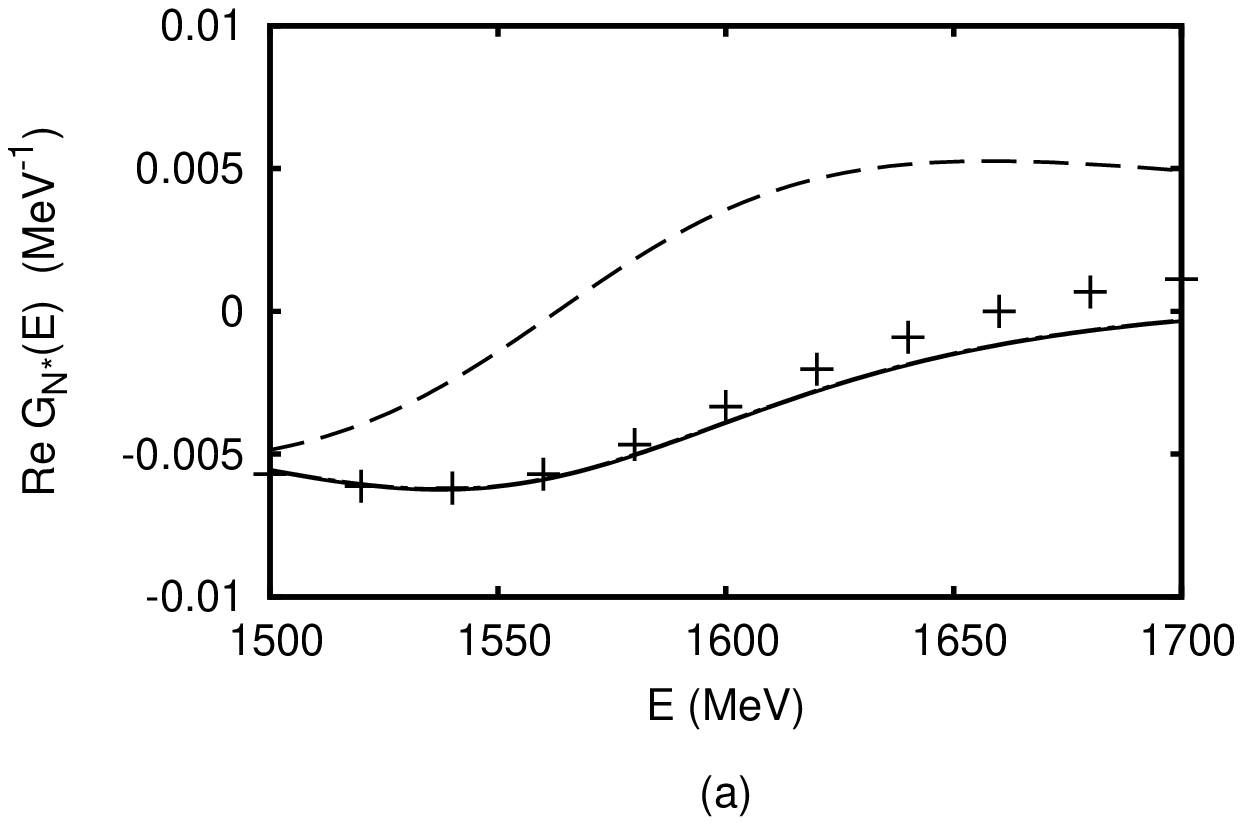}
\includegraphics[width=8cm]{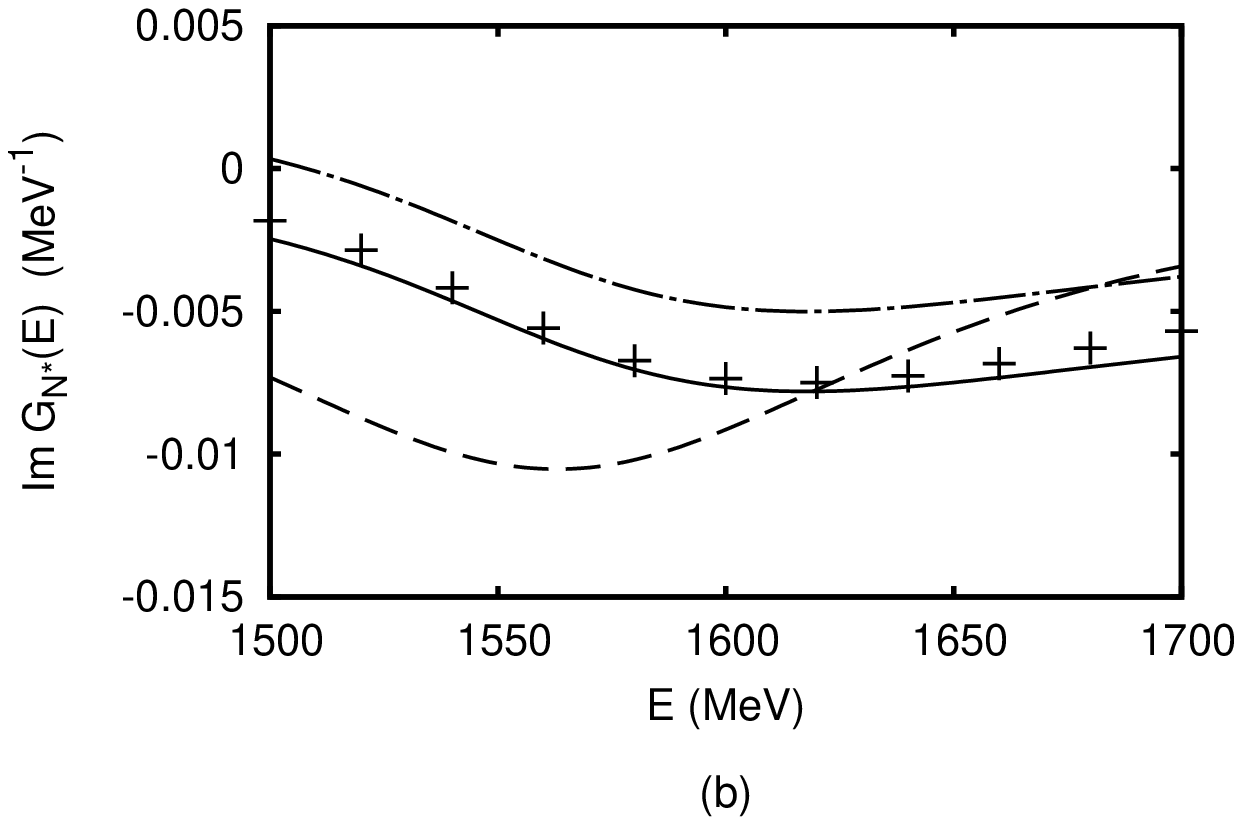}
\caption{ Comparisons of various resonance propagators: 
 exact propagator $G_{N^*}(E)$ (cross),
 $G_{N^*}^{(0)}(E)$ (dashed curve),
$G_{N^*}^{(1)}(E)$ (dash-dotted curve) 
 and $G_{N^*}^{(2)}(E)$ (solid curve). The real (imaginary) parts
are shown in the (a) ((b)) parts of the figure.
}
\label{fig:nstar-g}
\end{center}
\end{figure}

With the above expansion Eq. (\ref{eq:nstar-g}),
 we can introduce three different approximate forms for the $N^*$ propagator
\begin{eqnarray}
G_{N^*}^{(0)}(E)&=&\frac{1}{E-M},
\label{eq:natsr-g0} \\
G_{N^*}^{(1)}(E)&=&\frac{1}{E-M}\cdot\frac{1}{1-\Sigma'(M)},
\label{eq:nstar-g1}\\
G_{N^*}^{(2)}(E)&=&\frac{1}{E-M}\cdot\frac{1}{1-\Sigma'(M)}
+\frac{\Sigma''(M)}{2(1-\Sigma'(M))^2}.
\label{eq:nstar-g2}
\end{eqnarray}
In Fig.\ref{fig:nstar-g}, we compare the above three 
approximate propagators
with the exact result of $G_{N^*}(E)$ of Eq.  (\ref{eq:nstar-g-exact}).
The simple $G_{N^*}^{(0)}(E)$ (dashed curves) of Eq. (\ref{eq:natsr-g0})
are far from the  exact green function $G_{N^*}(E)$ (cross)
defined by Eq. (\ref{eq:nstar-g-exact}). When the factor
$1/(1-\Sigma'(M))$ is included, we obtain the dashed-dotted curves
for $ G_{N^*}^{(1)}(E)$. The solid curves are from 
$G_{N^*}^{(2)}(E)$. We see that
the constant term $\frac{\Sigma''(M)}{2(1-\Sigma'(M))^2}$
of Eq. (\ref{eq:nstar-g2}) mainly 
affects the imaginary part of the propagator.

\begin{figure}
\begin{center}
\includegraphics[width=8cm]{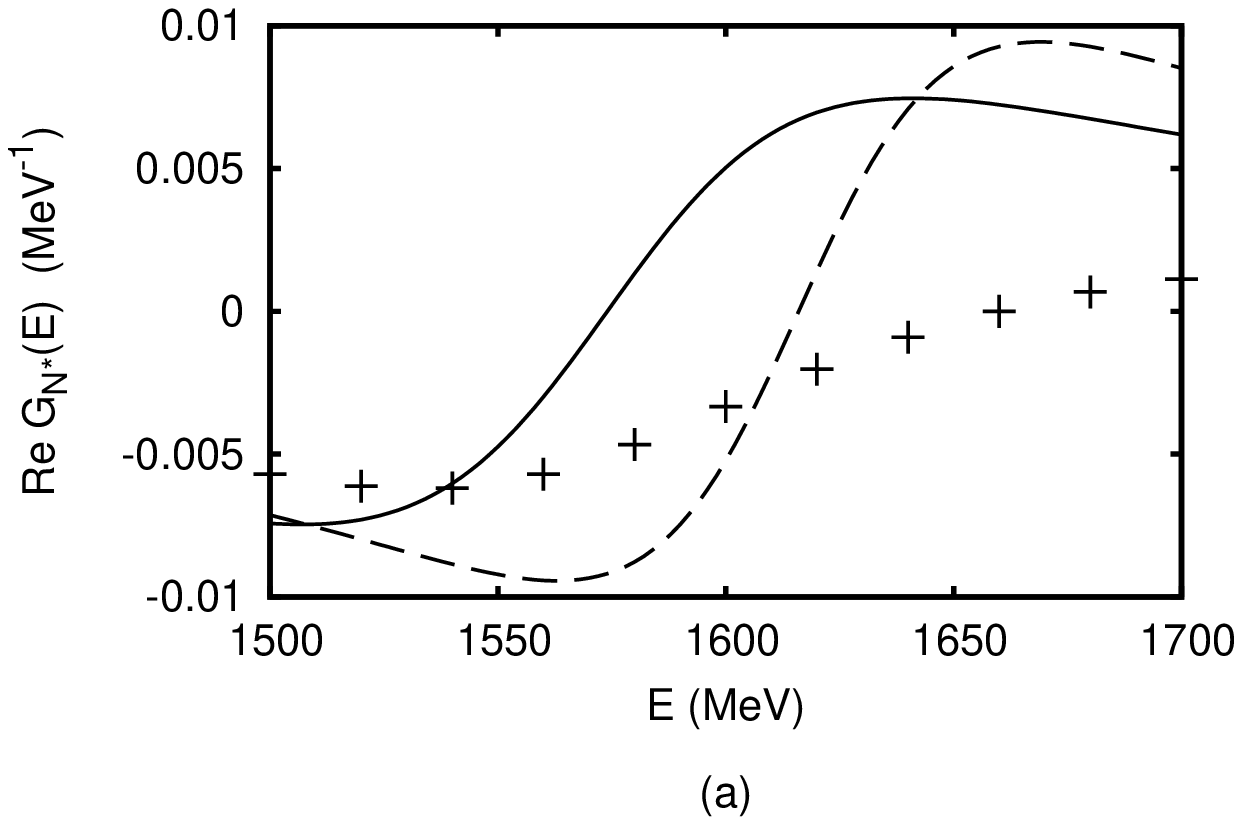}
\includegraphics[width=8cm]{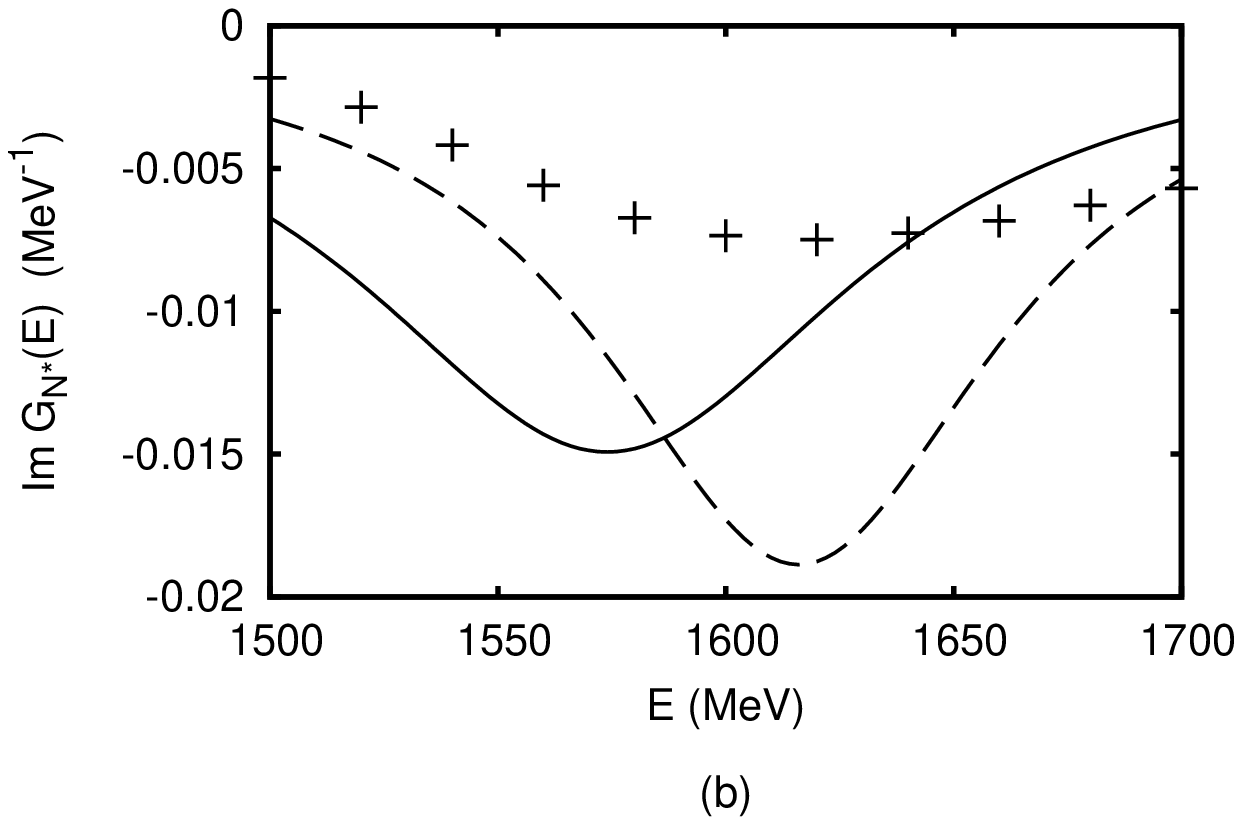}
\caption{The resonant propagator
 $G_{N^*}(E)$ (cross)
is compared with 
 $G_{N^*, SP}(E)$ (solid curve) using the SP poles and
$G_{N^*, TD}(E)$ (dashed curve) using the TD poles.
The real (imaginary) parts
are shown in the (a) ((b)) parts of the figure.
}
\label{fig:nstar-g1}
\end{center}
\end{figure}

Eq. (\ref{eq:nstar-g}) shows  that the simple pole approximation 
$G_{N^*}^{(0)}(E)=\frac{1}{E-M}$, which can be cast into the
usual Breit-Wigner form $1/(E-M_R + i\Gamma_R/2)$ with
$M_R=\mathrm{Re}\, M$ and
$\Gamma_R = -2\mathrm{Im}\, M$, works poorly. We also find that
the pole parametrization of the resonant propagator
using the poles extracted by  the TD and SP methods also work poorly.
For the considered $S_{31}$ case, the pole positions from using these two
methods are:
$E_{SP}=(1574,-67i)$ and $E_{TD}=(1616,-53i)$, as given in Table \ref{tab:ress11s31}.
In Fig.\ref{fig:nstar-g1}, we compare the exact propagator
(cross)
$G_{N^*}(E)$ of Eq. (\ref{eq:nstar-g-exact})
with the following two 
propagators
\begin{eqnarray}
G_{N^*, SP}(E) & = & \frac{1}{E-E_{SP}}, 
\label{eq:nstar-g-sp}\\
G_{N^*, TD}(E) & = & \frac{1}{E-E_{TD}}.
\label{eq:nstar-g-td}
\end{eqnarray}
 Clearly, phenomenological forms
Eqs. (\ref{eq:nstar-g-sp})-(\ref{eq:nstar-g-td}) can not account for the
complex coupled-channel resonant mechanisms.

\section{Summary}

In this paper, we have presented a pedagogical study of the commonly
used Speed-Plot (SP) and Time-Delay (TD)
methods for extracting the resonance parameters
from the empirically determined partial-wave amplitudes. Using a two-channel
Breit-Wigner form of the S-matrix, we show that
the poles extracted  by using theses two methods are 
on different Riemann sheets.
The SP method can find resonance poles on the unphysical $uu$-sheet, while
the TD method can find poles and zeros of S-matrix on $uu$ or $pp$-sheets
and therefore its validity is sensitive to the poles on $up$ or $pu$-sheets.
Furthermore, we also show numerically that these two
methods can fail to find those poles. Our results support the previous 
findings that
these two methods must be used with cautions in 
searching for  nucleon resonances from the meson-nucleon reaction data
in the region where the coupled-channel effects are important.  

We then develop an analytic continuation method for extracting the
resonance poles 
within a  Hamiltonian formulation of
meson-nucleon reactions.  
The main focus is on resolving the complications due to the coupling 
 with the unstable $\pi \Delta$, $\rho N$, and $\sigma N$ channels
which can decay into $\pi\pi N$ states.
Explicit numerical procedures are
presented and verified within several exactly solvable models.
The results from these models are also used to further demonstrate
the limitation of the SP and TD methods.

As a first application of the developed analytic continuation method, 
we present
the results from analyzing the $S_{11}$, $S_{31}$, $P_{33}$, $D_{13}$
and $F_{15}$ amplitudes of the dynamical coupled-channels model of $\pi N$
reactions developed
in Ref.\cite{jlms}. We also analyze the resonance
propagators and show that the simple
pole parametrization of the resonant propagator using the poles
extracted from SP and TD methods works poorly.

With the progress made in this work, we can proceed to extract 
all nucleon resonance parameters 
within the model of Ref.\cite{jlms}.
 However, this can be done more accurately only when 
the coupling with the unstable $\pi\Delta$, $\rho N$ and
$\sigma N$ channels are better determined by also fitting the
two-pion production data. Our progress in this direction will be 
reported elsewhere. 

\begin{acknowledgments}
 This work is supported by the Japan Society for the Promotion of Science,
Grant-in-Aid for Scientific Research(C) 20540270, and by the 
U.S. Department of Energy, Office of Nuclear Physics Division, under 
contract No. DE-AC02-06CH11357, and Contract No. DE-AC05-060R23177 
under which Jefferson Science Associates operates Jefferson Lab.

\end{acknowledgments}

\end{document}